
\input lanlmac

\def\LG{Lan\-dau-Ginz\-burg\ }

\def\half{{1 \over 2}}

\def\sk3{{\sqrt{k+3}}}
\def\half{{1 \over 2}}

\def\ssk2{{\sqrt{k+2}}}
\def\s3{{\sqrt{3}}}
\def\advm#1{{\it Adv. \ Math.}\ {\bf #1\/}}

\def\cmp#1{{\it Commun.\ Math. \ Phys.} \ {\bf #1\/}}
\def\nup#1{{\it Nucl.\ Phys.} \ {\bf B#1\/}}
\def\plt#1{{\it Phys.\ Lett.}\ {\bf #1\/}}
\def\ijmp#1{{\it Int. \ J.\ Mod. \ Phys.}\ {\bf A#1\/}}

\def\prd#1{{\it Phys.\ Rev.\ } {\bf D#1}\/}
\def\prpt#1{{\it Physics \ Reports\ } {\bf #1}\/}
\def\mpl#1{{\it Mod.\ Phys.\ Lett.} \   {\bf A#1}\ }

\def\coeff#1#2{\relax{\textstyle {#1 \over #2}}\displaystyle} 
\def\inbar{\vrule height1.5ex width.4pt depth0pt}
\def\IC{\relax\,\hbox{$\inbar\kern-.3em{\rm C}$}}
\def\IR{\relax{\rm I\kern-.18em R}}
\font\sanse=cmss12
\def\ZZ{\relax{\hbox{\sanse Z\kern-.42em Z}}}

\def\shalf{\coeff{1}{2}}

\noblackbox

%
\def\Titletwo#1#2#3#4{\nopagenumbers\abstractfont\hsize=\hstitle
\rightline{#1}\rightline{#2}
\vskip .7in\centerline{\titlefont #3}
\vskip .1in \centerline{\abstractfont {\titlefont #4}}
\abstractfont\vskip .5in\pageno=0}

\def\Date#1{\leftline{#1}\tenpoint\supereject\global\hsize=\hsbody%
\footline={\hss\tenrm\folio\hss}}
%


\Titletwo{}{}
{The Refined Elliptic Genus and Coulomb Gas Formulations
$^*$ { \abstractfont
\footnote{}{$^*$ Work supported in part by funds provided by the DOE
under grant No. DE-FG03-84ER40168.}} } { of $N=2$ Superconformal
Coset Models}
\centerline{D. Nemeschansky$^{**}$}
\footnote{}{$^{**}$ On leave of absence from the Physics Department,
University of Southern California, Los Angeles, CA 90089}
\centerline{Theory Division }
\centerline{CERN}
\centerline{Geneva, Switzerland}
\bigskip \centerline{ and }
\centerline{N.P. Warner}
\bigskip \centerline{Physics Department}
\centerline{University of Southern California}
\centerline{University Park}
\centerline{Los Angeles, CA 90089-0484.}
\vskip 1.0 cm

We describe, in some detail, a number of different Coulomb gas
formulations of $N=2$ superconformal coset models.  We also
give the mappings between these formulations.  The ultimate
purpose of this is to show how the Landau-Ginzburg structure
of these models can be used to extract the $W$-generators, and
to show how the computation of the elliptic genus can be refined
so as to extract very detailed information about the characters
of component parts of the model.

\vfill
\leftline{USC-94/018, \ \ CERN-TH.7548}
\leftline{hep-th/9412187}
\Date{December 1994}
%


\newsec{Introduction}

It has long been known that the $N=2$ superconformal coset models
based upon:
\eqn\coset{{\cal G}_n \ = \ {SU_k(n+1)\times SO(2n) \over
SU_{k+1}(n) \times U(1) } \ \ }
have \LG formulations and also have an underlying $N=2$
super-$W_{(n+1)}$ chiral algebra.  It is natural to ask if one can
determine, solely from a \LG formulation, whether or not the
corresponding $N=2$ superconformal model has such an extended
chiral algebra.  Moreover, given that a model has an $N=2$
super-$W$ algebra, one would like to know to what extent one can
determine
the spectrum of the zero-modes of the chiral algebra by using
the \LG structure alone.

\nref\WitLGa{E.~Witten, \nup{403} (1993) 159 .}
\nref \WitLGb{E.~Witten,``On the Landau-Ginzburg description of $N=2$
minimal
models,'' IASSNS-HEP-93-10, hep-th/9304026.}
\nref\PDFSY{P.~di Francesco and S.~Yankielowicz, \nup{409} (1993) 18;
P.~di Francesco, O.~Aharony and S.~Yankielowicz,
\nup{411} (1994) 584.}
\nref\Japan{ T.~Kawai, Y.~Yamada and S.-K.~Yang, ``Elliptic
genera and $N=2$ superconformal field theory,'' KEK-TH-362,
hep-th/9306096; T.~Kawai, ``Elliptic Genera  of $N=2$ hermitian
symmetric space models,'' preprint KEK-TH-403, hep-th/9407009;
T.~ Kawai and S.-K.~ Yang, ``Duality of Orbifoldized Elliptic
Genera,'' preprint KEK-TH-409, hep-th/9408121.}
\nref\MHenn{ M.~Henningson, ``$N=2$ gauged WZW models and the
elliptic genus,'' IASSNS-HEP-93-39, hep-th/9307040;
P.~Berglund and M.~Henningson, ``\LG orbifolds, mirror symmetry
and the elliptic genus,'' IASSNS-HEP-93-92, hep-th/9401029;
``On the Elliptic Genus and Mirror Symmetry,''
IASSNS-HEP-94-11, hep-th/9406045.}

Techniques by which one can answer these questions were introduced in
\refs{\WitLGa, \WitLGb}.  In particular, a number of the
Ramond sector characters can be extracted from the elliptic genus,
and it was shown how the latter can easily be calculated
from the \LG formulation.
Similar computations of Ramond characters, but refined by $N=2$,
$U(1)$
charge, were performed for more complex models in \refs{\PDFSY
-\MHenn}.
However for models with central charge $c \ge 3$, the elliptic genus,
even when refined by the $U(1)$ charge, is too coarse to determine
the
complete structure of the Hilbert spaces constructed above the Ramond
ground
states.  To completely characterize these Hilbert spaces, one needs
to look
for extended chiral algebras and then appropriately refine the
elliptic genus as in \ref\DNNWgen{D.~Nemeschansky and N.P.~Warner,
\plt{B329} (1994) 53.}.

\nref\WLAS{W.~Lerche and A.~Sevrin, ``On the Landau-Ginzburg
Realization of Topological Gravities,'' CERN preprint TH-7210/94,
hep-th/9403183.}

Using methods of \WitLGb, it was shown in \ref\KMohri{K.~Mohri,
``N=2 Super-$W$ algebra in half-twisted
Landau-Ginzburg model'', UTHEP-260, hep-th/9307029.}
that at the classical level, the form of the superpotential
determines
whether there is a super-$W$ algebra acting upon the elliptic
cohomology of the theory.   Quantum versions of these results were
obtained in \refs{\DNNWgen,\WLAS}, and in \DNNWgen\ it was also shown
how the elliptic genus could be refined to yield much more complete
information about the structure of the Ramond sector Hilbert space.

Our primary purpose in this paper is to expand and develop the
results of our earlier letter \DNNWgen. In addition to doing this we
also wish to discuss the relationships between the many fomulations
of the coset model \coset. This will be done in section 3. In doing
this we will encounter an interesting feature of Coulomb gas
formulations tensor products of conformal models with a special
choice of modular invariant. Partly for its own interest, and partly
as preparation for section 3, we will exhibit a simple example of
this feature in section 2. Indeed, section 2 can be read
independently of the rest of the paper.

In section 4, we will descibe, as simply as possible, the $N=2$
super-$W$ algebra by giving an explicit method for constructing the
lowest components of each of the chiral algebra superfields. We then
re-express this in terms of \LG fields and show how it can be used to
refine the elliptic genus completely with respect to the super-$W$
algebra. In section 5 we expand the fully refined elliptic genus and
obtain formulae for the branching functions that make up the model.
Finally, we discuss some issues about fermionic screening currents in
the Coulomb gas formulation, and indicate how this might possibly be
used to extract information about the modular invariant partition
function of the complete model.

\newsec{Coulomb gas formulations of special tensor product models}

The  model that we wish to consider in this section is the coset
theory
${SU(2)_k\times SU(2)_1\times SU(2)_1\over SU(2)_{k+2}}$.
This can be written in terms of a tensor product of minimal models:
\eqn\decotoy{ {\cal M}_1 \times {\cal M}_2  \ =\
{ SU(2)_k \times SU(2)_1 \over SU(2)_{k+1} } \times { SU(2)_{k+1}
\times SU(2)_1 \over SU(2)_{k+2} } \ \ . }
However, one has to remember that to recover the original model, the
reperesentation of the denominator factor of $SU_{k+1}(2)$ in ${\cal
M}_1$ must always be the same as that of the numerator factor of
$SU_{k+1}(2)$ in ${\cal M}_2$. This ``locking together'' of
representations defines a special modular invariant of the tensor
product model.

The  stress-tensors for the Coulomb gas formulation of
${\cal M}_1$ and ${\cal M}_2 $ are
\eqn\stress{\eqalign{ T_1(z) \ = & \ -\coeff{1}{2}(\del \phi_1 )^2 +
i(\alpha_+ - \alpha_-) \del^2\phi_1 \cr
T_2(z) \ = & \ -\coeff{1}{2}(\del \phi_2 )^2 +i(\beta_+ - \beta_-)
\del^2\phi_2  \ \ , \cr  }}
where we define:
\eqn\alphap{ \alpha_{\pm} \ = \ \Bigl ( {k+2\over k+3} \Bigr )^
{\pm \coeff{1}{2} } \qquad {\rm and} \qquad \beta_{\pm} \ = \
\Bigl ( {k+3\over k+4} \Bigr )^{\pm \coeff{1}{2} } \ .}
The primary fields in the two models can be represented in terms of
vertex operators:
\eqn\primaryv{\eqalign{V^{(1)}_{m,n} \ = & \ e^{-{i \over \sqrt 2}
(m\alpha_+- n \alpha_-) \phi_1 } \ , \cr
V^{(2)}_{m,n} \ = & \ e^{-{i \over \sqrt 2}(m\beta_+-
n \beta_-) \phi_2 }\ , \  }}
whose conformal dimensions are given by:
\eqn\condim{\eqalign{\Delta^{(1)}_{m,n}\ = & \
\coeff{1}{4}[ (m+1)\alpha_+ - (n+1) \alpha_-]^2 - \coeff{1}{4}
(\alpha_+-\alpha_-)^2 \ , \cr
\Delta^{(2)}_{m,n}\ = & \ \coeff{1}{4}
[ (m+1)\beta_+ - (n+1) \beta_-]^2-\coeff{1}{4} (\beta_+-\beta_-)^2
\ \ .\cr }}
In particular, the vertex operators:
\eqn\dualvac{W^{(i)} ~\equiv ~ V^{(i)}_{-2, -2} }
are the dual representatives of the vacuum states.
The screening currents in  ${\cal M}_1$ and ${\cal M}_2$ have the
form
\eqn\scretwo{R^{\pm} \ = \  e^{\pm i \sqrt 2 \alpha_{\pm} \phi_1 }
\qquad {\rm and} \qquad S^{\pm} \ = \  e^{\pm i \sqrt 2
\beta_{\pm} \phi_2 }  \  .  }

The tensor product of the minimal models can be constructed using the
screening currents $R^\pm$ and $S^\pm$ independently.  The tensor
product
then has an obvious spin-$2$ element of the chiral algebra:
\eqn\simpleS{S(z) ~=~ c_2 T_1(z) ~-~ c_1 T_2(z) \ . }
The coefficients, $c_1$ and $c_2$, are the central charges of the
two minimal models, and the foregoing combination of $T_1$ and $T_2$
is a good conformal field with respect to the total stress-tensor.

The locking together of representations of $SU_{k+1}(2)$ in the
tensor
product means that the allowed vertex operators have the form:
\eqn\lockedvert{V_{m,n,p} ~\equiv~ V_{m,n}^{(1)} ~ V^{(2)}_{p,m}
{}~=~ e^{{i \over \sqrt{2}} [(n\alpha_- \phi_1 ~-~ p \beta_+ \phi_2)
{}~-~
m(\alpha_+ \phi_1 ~-~ \beta_- \phi_2)]} \ .}
Consider the operators:
\eqn\Xdefn{{\cal X}_r ~\equiv~ V^{(1)}_{-2r,0} ~  V^{(2)}_{0,-2r}
{}~=~ e^{i \sqrt{2}   r (\alpha_+ \phi_1 ~-~ \beta_- \phi_2)} \ .}
Note that ${\cal X}_1 = R^+ S^-$. The operator ${\cal X}_r$ has
dimension $2r^2$ and is local with respect to all of the vertex
operators $V_{m,n,p}$. Thus the operators ${\cal X}_r$ can be thought
of as elements of an extended chiral algebra in the free bosonic
theory. They are also local with respect to $R^-$ and $S^+$, but not
with respect to $R^+$ and $S^-$.

Generally the operators
${\cal X}_r$ do not play any role after one has reduced to the
simple tensor product of minimal conformal models.
However, if one follows the spirit of locking the $SU_{k+1}(2)$
representations together, then it is much more natural to
introduce the screening charges:
\eqn\modscreen{\eqalign{
Q_+^{(1)} \ =\ & \oint V_{-2,0,-2} \ =\  \oint R^+ W^{(2)} \ = \
\oint  e^{i \sqrt 2 \alpha_+ \phi _1}
e^{i \sqrt 2 (\beta_+ -\beta_-) \phi_2} \cr
Q_-^{(1)} \ =\ & \oint V_{0,-2,0} \ =\ \oint R^-
 \ = \ \oint e^{-i \sqrt 2 \alpha_- \phi _1}
\cr
Q_+^{(2)} \ =\ & \oint V_{0,0,-2} \ =\  \oint   S^+
 \ = \ \oint e^{i \sqrt 2 \beta_+ \phi _2}
\cr
Q_-^{(2)} \ =\ & \oint V_{-2,-2,0} \ =\  \oint  W^{(1)}S^- \ = \
\oint e^{i \sqrt 2 (\alpha_+ -\alpha_-) \phi_1}
 e^{-i \sqrt 2 \beta_- \phi _2}  \cr
}}

Observe that the current in $Q_+^{(1)}$ is simply $R^+$ multiplied by
the
dual vacuum vector of ${\cal M}_2$, and similarly for $Q_-^{(2)}$.
One then finds that \simpleS\ no longer commutes with the screening
charges.  However, define
\eqn\repS{S \ = \ c_2 T_1 ~-~ c_1 T_2 ~+~ \xi R^+ S^- \ \ , }
for some constant $\xi$.  One then finds
\eqn\truescre{\eqalign{[~Q_-^{(1)} ~+~ \zeta ~ Q^{(2)}_- ~,~ S~ ] \ =
\ & 0
 \cr [~\zeta ~Q_+^{(1)} ~+~ Q^{(2)}_+ ~,~ S~]\ = \ & 0 \ \ , \cr }}
where  $\zeta \ = \ {\beta_+\xi\over (c_1+c_2)(\beta_+-\beta_-)}$
$ = {\alpha_-\xi\over (c_1+c_2)(\alpha_+-\alpha_-)}$.
Thus in the locked model the screening currents involve the dual
vacuum
vectors, and the naive representations of non-trivial elements of the
chiral algebra receive nilpotent vertex operator corrections.

In the next section we will encounter  examples of the foregoing
``locked'' tensor product model.  Moreover, they naturally come
equipped
with screening charges analogous to \modscreen.  We will find it
convenient to convert this description into the simple version of the
tensor product where one has to remember to lock the representations,
but in which the screening currents are not mixed with dual vacua,
and in which the chiral algebra contains no
nilpotent vertex operator parts.

\newsec{The multifarious formulations of the $N=2$
super-$W_{n+1}$ models}

\nref\LVW{W.~Lerche, C.~Vafa and N.P.~Warner, \nup{324} (1989) 427.}
\nref\Gepner{D.~Gepner, \nup{322B} (1989) 65.}
\nref\Fre{P.~ Fr\'e, L.~Girardello A.~ Lerda and P.~Soriani
\nup{387} (1992) 333.}
\nref\DNSY{D.~Nemeschansky and S.~Yankielowicz,``$N=2$ W-algebras,
Kazama-Suzuki Models and Drinfeld-Sokolov Reduction'', USC-preprint
USC-007-91 (1991).}
\nref\KIto{K.~Ito, \plt{259} (1991) 73.}

The first and most obvious formulation of the model \coset\ is as a
coset model \ref\KS{Y.~Kazama and H.~Suzuki, \nup{234} (1989) 73.}.
There is also a formulation that comes from Drinfel'd-Sokolov
reduction \refs{\DNSY,\KIto}.
There is the \LG formulation \refs{\LVW,\Gepner}, and a related
Coulomb gas description \refs{\Fre,\WitLGb}. There is also a Coulomb
gas description of the model considered as a tensor product. We will
consider all of these formulations here, and describe how they are
related. We begin with the coset formulation.

\subsec{The structure of the coset model}

We will not review the details of \KS, but simply wish to
describe some of the general structure of the that can be seen
from the coset formulation.

We begin by observing that the model \coset\ can be written as a
tensor
product:
\eqn\tensor{\eqalign{ {SU_k(n+1) \times SO(2n) \over
SU_{k+1}(n) \times U(1) } & ~=~ {\cal M}_1 ~\times~ {\cal M}_2
{}~\times~ U(1) \cr
& ~=~ {SU_k (n+1) \over SU_k(n) \times U(1)} ~\times~ {SU_k (n)
\times SU_1(n)  \over SU_{k+1} (n)} ~\times~ U(1)  \ ,} }
where ${\cal M}_2$ is a non-supersymmetric $W_n$-model.  However,
exactly as in the last section, the representations of $SU_k(n)$
in ${\cal M}_1$ and ${\cal M}_2$ are locked together.

The chiral algebra of the $N=2$ supersymmetric model contains a
non-supersymmetric $W_n$ subalgebra.  We now wish to argue that for
$k$
sufficiently large, the generators of this $W_n$, along with the
$U(1)$
current, provide a complete set of lowest spin (bottom) components
of the $N=2$ superfields that make up the full $N=2$ chiral algebra.
In doing this we will also elucidate a duality between the chiral
algebra
and chiral ring of \coset.  This has been well known for some time
\ref\lNW{W.~Lerche, D.~Nemeschansky and N.P.~Warner, unpublished},
and the classical version of it has been described in \KMohri.

\nref\EY{R.~Sasaki and I.~Yamanaka, in
``Conformal Field Theory and Solvable Models,''
{\it Advanced Studies in  Pure Mathematics } {\bf 16}, 1988.
T.~Eguchi and S-K. Yang,\plt{224B} (1989) 373;
\mpl{5} (1990) 1693. }
\nref\BilGe{A.~Bilal and  J.-L. \ Gervais, \plt{206} (1988) 412;
\nup{318} (1989) 579; \nup{326} (1989) 222;
A.\ Bilal, \nup{330} (1990) 399;  \ijmp{5} (1990) 1881.}
\nref\HM{T.~Hollowood and P.~Mansfield, \plt {226} (1989) 73.}
\nref\FatL{V.~Fateev and V.~Lukyanov, Int. J. Mod. Phys. {\bf A3}
(1988) 507. }
\nref\FLMW{P.~Fendley, W.~Lerche,
S.D.~Mathur, and N.P.~Warner, \nup{348} (1991) 66.}
\nref\WLNW{W.~Lerche and N.P.~Warner, \nup{358} (1991) 571.}

The first thing to observe is that, for $k$ large, there are $n$
independent supermultiplets in the $N=2$ chiral algebra, and that the
spins of the lowest components are $1,\dots,n$. This matches the
spins of the generators of the $W_n \times U(1)$ algebra. It also
matches the $N=2$, $U(1)$ charges of the generators of the chiral
ring. Next, one considers an equivalent formulation of the coset
model \KS:
\eqn\otherform{\eqalign{ {G \times SO(dim(G/H) \over H} &~\equiv~
{SU_1(k+n+1) \times SO(2kn) \over SU_{n+1} (k) \times SU_{k+1} (n)
\times U(1) } \cr
& ~=~  {SU_1 (k) \times SU_n (k) \over SU_{n+1} (k)} ~\times~
{SU_1 (n) \times SU_k (n) \over SU_{k+1} (n)} ~\times~ U(1)  \ .} }
Note that the second factor is ${\cal M}_2$.  One can define this
model
entirely in terms of free bosons \FLMW.  The elements of the
chiral algebra can be represented by those polynomials in derivatives
of the bosons that are invariant (up to total derivatives)  under the
Weyl group, $W(H_0)$, of $H_0 = SU (k) \times SU (n)$
\refs{\EY {--} \FLMW}.  In \FLMW\ it was shown that the
supercurrents
could be represented by vertex operators that are related to
screening
currents via the action of the
maximal cyclic generator of the Weyl group,
$W(G)$, of $G$.  Thus the top components of superfields are those
polynomials in derivatives of bosons that are invariant (up to total
derivatives) under the action of $W(G)$.  This is because $W(G)$
invariance
means that the polynomial will commute (up to total derivatives) with
the
supercharges.

There is now a natural finite
ring structure that we can define upon the chiral algebra:  consider
all the $W(H_0)$ invariant polynomials, modulo the $W(G)$ invariant
polynomials.  This set consists of all the polynomials in elements of
the chiral algebra such that these polynomials
are not top components of superfields.   This is also the
characterization, in terms of `Cartan subalgebra variables,'  of the
chiral
ring of the $N=2$ superconformal model \refs{\LVW,\WLNW}.
It is also a well know fact that the foregoing ring can be
generated by restricting to the bosons that correspond to either the
first or second factors in \otherform. (This fact was used in \LVW\
to
construct the \LG potential for the model.)  Thus, this finite
quotient
ring of the chiral algebra is isomorphic to the chiral ring.
Moreover,
the ring can be generated by the chiral algebra generators of the
second factor of \otherform, that is, by the $W_n$ algebra of
${\cal M}_2$.  Therefore, the task of finding representatives of the
$N=2$
superconformal chiral algebra is complete once we have the
supercharges
and either the $W_n$ algebra, or some free bosonic realization of
${\cal M}_2$.

The fact that a description of the chiral ring can be mapped onto the
foregoing quotient ring of the chiral algebra will not be important
to this paper, and we have included merely for interest's sake.
We feel that one should be able to establish this relationship
more directly within the superconformal model itself,
and that one should be able to use it to
understand the conserved charge structure discussed in \WLNW\ for
solitons of the quantum integrable, off-critical models based upon
\coset.

\subsec{Supersymmetric Drinfel'd -- Sokolov reduction}

The free  superfield formulation of \coset\  can be
obtained from the Lie superalgebra $A(n,n-1)$ through a Hamiltonian
reduction \refs{\DNSY,\KIto}.
Before describing the free field formulation we first review
some basic properties of the super Lie algebra $A(n,n-1)$ that are
relevant to our discussion \ref\supKac{V.G. \ Kac,
{\it Adv. Math. } {\bf 26} (1977) 8.}.

The algebra $A(n,n-1)$ has a $\ZZ_2$-grading under which roots are
viewed as either even or odd.  If we denote the simple roots by
$\alpha_1,\alpha_2,\ldots,\alpha_{2n-1}, \alpha_{2n} $, then the even
roots are:
\eqn\evenroots{\alpha_i +  \alpha_{i+1} + \ldots +
\alpha_{i+2k-2} + \alpha_{i+2k-1} \qquad  k=1,2, \ldots ,\Bigl
[{2n+1-i \over 2 } \Bigr ] \  ,}
and the odd roots are:
\eqn\oddroots{\alpha_i + \alpha_{i+1} + \ldots + \alpha_{i+2k-1} +
\alpha_{i+2k} \qquad k =0,1,\ldots , \Bigl [{2n-i \over 2 } \Bigr
] \ . }
The simple roots of $A(n,n-1)$ satisfy the following
relations:
\eqn\rootrel{\alpha_{2i-1} \cdot \alpha_{2i} \  = \ 1 \ ; \qquad
 \alpha_{2i+1} \cdot \alpha_{2i} \ = \ -1  \ . }
All other inner products are zero (including $\alpha_i \cdot
\alpha_i$).  The fundamental weights $\lambda_1, \ldots ,
\lambda_{2n}$
are defined by:
\eqn\funweig{ \alpha_i \cdot \lambda_j \ = \ \delta_{ij}   \ \ . }
It is easy to see from \rootrel,
that in terms of the simple roots, the fundamental weights are given
by:
\eqn\weightroot{\eqalign{ \lambda_{2i} \ & = \ \alpha_1 + \alpha_3
+ \ldots + \alpha_{2i-3} + \alpha_{2i-1} \ , \cr \lambda_{2i-1} \ & =
\ \alpha_{2i} + \alpha_{2i+2} + \ldots \alpha_{2n-2}+ \alpha_{2n}  \
{}.
\cr }}

The super Lie algebra $A(n,n-1)$ contains the even subalgebras
$A_n$ and $A_{n-1}$.  The simple roots of these two subalgebras are
given respectively by:
\eqn\evensimple{ \alpha_{2i-1} + \alpha_{2i} \ , \quad i=1,\ldots ,n
;
\quad {\rm and} \quad \alpha_{2i} + \alpha_{2i+1} \ ,
\quad   i=1, \ldots, n-1 \  .}
{}From \rootrel\ we see that  the root system for $A_n$ has a
positive
definite metric, whereas for $A_{n-1}$, the metric is negative
definite.

To write down the free superfield description of \coset\ it is most
convenient to use an $N=1$ superfield formulation.  We therefore
introduce a  single anti-commuting coordinate $\theta$, and define
the
super-derivative, $D$ by:
\eqn\superder{ D \ = \ {\partial \over \partial
\theta } + \theta {\partial \over {\partial z}} \ .}
Consider $2n$ (real) superfields
\eqn\superfield{ \Phi^i(z,\theta) \ = \ \varphi^i(z) + \theta
{}~\chi^i(z)
\ , }
where $\varphi^i(z)$ is a free bosonic field and $\chi^i(z)$ is
a free, real fermion.  These superfields satisfy the operator
product expansion:
\eqn\susyope{\Phi^i(z_1,\theta_1)~
\Phi^j(z_2,\theta_2) \ = \ - ~\delta^{ij}~log (z_{12}) \ ,}
where  $z_{12} \ \equiv \ z_1-z_2 -\theta_1 \theta_2$.
In terms of components, we have:
\eqn\compOPEs{\varphi_i(z)~\varphi_j(w) ~=~ - ~\delta^{ij} ~ log(z-w)
\ ,
\qquad \chi_i(z)~\chi_j(w) ~=~ - ~ \delta^{ij} ~ {1 \over (z-w)} \ .}

The generators of the extended chiral algebra are then obtained from
the
Lax operator \refs{\DNSY, \KIto}:
\eqn\laxope{L \ = \ \prod_{j=1}^{2n+1} \big [~ i \alpha_0 D ~-~
(-1)^j
(\lambda_j - \lambda_{j-1}) \cdot D\Phi~ \big] \ \ , }
where  $\lambda_0 \equiv \lambda_{2n+1} \equiv \ 0$.

The parameter $\alpha_0$ is background charge of Feigin-Fuchs
representation.  In order to reproduce \coset, whose central charge
is
$ c = {3k n \over k+n+1}$,  we must  set
\eqn\alphazero{ \alpha_0 \ = \ {1 \over {\sqrt{ k+n+1}}}  \  . }

In the $N=1$ superfield formulation the stress tensor $T(z)$ is the
top
component of an $N=1$ superfield $T(z,\theta)$  with conformal
dimension $3/2$,
\eqn\susyt{ T(z,\theta) \ = \ \shalf \big( ~G^+(z)
{}~+~  G^-(z)~\big) ~+~ \theta ~T(z) \  .}
The fields $G^\pm(z)$ in
\susyt\ are the two supersymmetry generators of the $N=2$
supersymmetry
algebra.  The $U(1)$ current, $J(z)$, of the $N=2$ algebra is the
lowest component of the superfield $J(z,\theta)$
\eqn\jone{J(z,\theta) \ = \ J(z) ~+~ \theta~ \shalf \big(~G^+(z) ~-~
G^-(z)~\big) \  . }

The free field forms of these superfields are obtained from the
quadratic and linear parts of the Lax operator.  One finds:
\eqn\deftz{\eqalign{ T(z,\theta) \ = &\ -\shalf~
\sum_{i=1}^{n} \lambda_{2i} \cdot D\Phi^i \alpha_{2i} \cdot \partial
\Phi^i ~-~ \shalf ~ \sum_{i=1}^{n} \alpha_{2i} \cdot D\Phi^i
\lambda_{2i}
\cdot \partial \Phi^i \cr & \ -{i \over 2 \sqrt{k+n+1}}
{}~\sum_{i=1}^{2n}
\lambda_i \cdot D^3 \Phi \ . }}
and
\eqn\defuones{J(z,\theta) \ =\ \sum_{i=1}^n
\big( \lambda_{2i}\cdot D\Phi \big) \big(\alpha_{2i}
\cdot D\Phi\big) ~-~{i \over \sqrt{k+n+1}}~ \sum_{i=1}^n
(\lambda_{2i}-
\lambda_{2i-1} ) \cdot \partial \Phi \ \ . }

To define the conformal model fully, we need the screening operators.
These are in one-to-one correspondence with the roots of the Lie
superalgebra $A(n,n-1)$ and its even subalgebras $A_n$ and $A_{n-1}$.
The screening operators corresponding to the roots of $A_n$ have the
form:
\eqn\anscren{ Q_{\alpha_{2i-1}+\alpha_{2i}} \ = \ \oint dz
d\theta \ (\alpha_{2i}-\alpha_{2i-1}) \cdot D\Phi ~ e^{-{ i \over
\sqrt{k+n+1}}(\alpha_{2i-1}+ \alpha_{2i} )\cdot \Phi} \ ,}
while the screening operators corresponding to the
roots of $A_{n-1}$ have the form:
\eqn\screanone{Q_{\alpha_{2i}+\alpha_{2i+1}} \ = \ \oint dz d\theta \
(\alpha_{2i} - \alpha_{2i+1} )\cdot D\Phi ~ e^{{+{i\over
\sqrt{k+n+1}}}(\alpha_{2i}+ \alpha_{2i+1} )\cdot \Phi} \ .}
These screening operators are usually called $D$-type screeners.
The screening operators associated the simple root of $A(n,n-1)$, are
usually referred to as $F$-type, or ``fermionic,'' screening
operators,
and these have the form:
\eqn\screenann{Q_{\alpha_i} \ = \ \oint dz
d\theta ~ e^{{i  \sqrt{k+n+1}} ~ \alpha_i\cdot \Phi} \ . }

It is relatively easy
re-express the foregoing in a manifestly $N=2$ supersymmetric
formalism.  We will adopt the following $N=2$ superfield conventions.
First introduce
\eqn\defD{D^\pm \ = \  {\del \over \del \theta^\mp} + \theta^\pm {
\del \over \del z } \qquad \bar D^\pm \ = \  {\del \over \del
\bar \theta^\mp} + \bar \theta^\pm { \del \over \del \bar z }  \ .}
These satisfy
\eqn\Drel{ \{ D^+,D^-\} \ = \ 2 ~{ \del \over \del z} \ ; \qquad
\{ \bar D^+,\bar D^-\} \ = \ 2 ~{ \del \over \del\bar z} \ .}
Let $\Phi_i^+(z, \theta^+,\theta^-)$ denote  a set of
$n$  holomorphic, chiral bosonic superfields. That is, they satisfy
\eqn\condi{ D^- ~ \Phi_i^+ \ = \ 0 \ ; \qquad \bar D^{\pm}~\Phi_i^+
\ = \ 0 \ . }
Similarly, $\Phi_i^-$  will denote conjugate anti-chiral bosonic
superfields, that satisfy
\eqn\theother{ D^+~\Phi_i^- \ = \ 0 \ ; \qquad \bar D^{\pm}~\Phi_i^-
\ = \ 0 \ .}
In terms of components, $\Phi_j^\pm$ can expanded as follows:
\eqn\comp{\eqalign{&\Phi_j^+(z,\theta^+,\theta^-) \ = \ \phi_j(z) ~+~
\sqrt{2} ~ \theta^- ~ \psi_j(z) ~+~ \theta^-\theta^+ ~\del \phi_j(z)
\cr
&\Phi_j^-(z,\theta^+,\theta^-) \ = \ \bar\phi_j(z) ~+~ \sqrt{2} ~
\theta^+
{}~\bar \psi_j(z) ~-~ \theta^-\theta^+ ~ \del \bar \phi_j(z) \ .  \cr
  }}
We take the operator product to be
\eqn\opeex{\Phi_i^{\pm}(z_1, \theta_1^+,\theta_1^-) ~
\Phi_j^{\mp}(z_2,
\theta_2^+,\theta_2^-)  \ \sim \ -~\delta_{ij}~ log(\tilde z_{12}
{}~\pm ~
\theta_{12}^-\theta_{12}^+)  \ , }
where $\theta_{12} = \theta_1-\theta_2$ and $\tilde z_{12}=
z_1-z_2-\theta_{1}^+\theta^-_{2}-\theta_1^-\theta_2^+$.

For the component fields this means that
\eqn\opecomp{\eqalign{ \phi_i(z_1) ~\bar \phi_j(z_2) \ & \sim \
-~\delta_{ij}~log(z_1-z_2) \ , \cr
\psi_i(z_1)~ \bar \psi_j(z_2) \ & \sim \ -~\delta_{ij}~
{1 \over z_1-z_2 } \ . \cr }}

To relate these to the $N=1$ superfields, we first note that
the weight space of $A(n,n-1)$ has a natural complex basis
spanned by $\alpha_{2j}$ and $\lambda_{2j}$, $j =1, \dots,n$.
These vectors satisfy:
\eqn\newbasis{\alpha_{2i}\cdot \alpha_{2j} ~=~ 0 \ ; \quad
\lambda_{2i}
\cdot  \lambda_{2j} ~=~ 0\ ; \quad \alpha_{2i} \cdot \lambda_{2j} ~=~
\delta_{ij} \ .}
We can then identify the complex components of the $N=2$ superfields
with the $N=1$ components according to:
\eqn\twoasone{\eqalign{ \phi_j &~=~ \lambda_{2j} \cdot \varphi\ ;
\qquad \bar \phi_j ~=~  \alpha_{2j}  \cdot \varphi \ , \cr
\psi_j &~=~ \lambda_{2j} \cdot \chi\ ; \qquad \bar \psi_j ~=~
\alpha_{2j}  \cdot \chi \ .} }

With these superfields, and using \defuones\ and \weightroot, one can
write
the complete energy momentum tensor as:
\eqn\emtensor{{\cal J} \  = \ + \coeff{1}{4}~ \sum_{j=1}^n ~ (D^+
\Phi^+_j)
{}~(D^-\Phi^-_j)  ~-~ i \ {\alpha_0 \over 2} \sum_{j=1}^n  \big[
{}~\del \Phi^+_j
{}~-~ j ~\del \Phi^-_j ~ \big ] \ .}

The components of ${\cal J }$ are simply the $N=2$ superconformal
generators, and they can be read off from the expansion:
\eqn\wthrsup{{\cal J} \  = \ \shalf~J  ~+~ \coeff{1}{\sqrt{2}}~
 \theta^+ G^+ ~-~ \coeff{1}{\sqrt{2}} ~ \theta^-G^-  ~+~
 \theta^- \theta^+ T \ . }
Explicitly, one has:
\eqn\ntwogen{ \eqalign{J(z)\ = \ & ~-~ \sum_{j=1}^n \big[
\bar \psi_j(z)~\psi_j(z) ~+~ i \alpha_0 \del \phi_j(z) - i j \alpha_0
\del \bar \phi_j(z) \big ]  \cr
G^{+}(z) \ = \ &  \sum_{j=1}^n \big[  \bar \psi_j(z) \del \phi_j(z)
{}~+~ i  j \alpha_0 ~ \del\bar \psi_j(z) \big ]   \cr
 G^{-}(z) \ = \ &  \sum_{j=1}^n \big[  \psi_j(z) \del \bar
\phi_j(z) ~+~
i \alpha_0 ~ \del \psi_j(z)  \big ] \cr
T(z) \ = \ & - \sum_{j=1}^n \big[( \del\phi_j) (\del\bar\phi_j) ~-~
\coeff{1}{2}(\psi_j \del\bar \psi_j + \bar \psi_j \del\psi_j)
{}~+~ i {\alpha_0 \over 2} \del^2 \phi_j  ~+~ i j {\alpha_0 \over 2}
\del^2 \bar \phi_j \big ] \ . \cr }}
The screening operators may also be similarly translated:
\eqn\newscreener{\eqalign{Q_{\alpha_{2i-1} + \alpha_{2i}} \ &= \
\oint dz
\ \big[ (\del \bar \phi_i - \del \phi_i + \del \phi_{i-1}) ~+~ \cr
& \qquad \qquad \qquad \qquad \qquad
2 i \alpha_0 (\bar \psi_i \psi_i - \bar \psi_i \psi_{i-1}) \big]
 ~ e^{- i \alpha_0 (\bar \phi_i + \phi_i -  \phi_{i-1}) } \ , \cr
Q_{\alpha_{2i}+\alpha_{2i+1}} \ &= \ \oint dz
\ \big[ (\del \bar \phi_i + \del \phi_i - \del \phi_{i+1}) ~+~ \cr
& \qquad \qquad \qquad \qquad \qquad
2 i \alpha_0 (\bar \psi_i \psi_i - \bar \psi_i \psi_{i+1}) \big]
 ~ e^{+i \alpha_0 (\bar \phi_i - \phi_i + \phi_{i+1} )}  \ , \cr
Q_{\alpha_{2i}} \ &= \ \oint dz~
\bar \psi_i  ~ e^{{i  \sqrt{k+n+1}} ~ \bar \phi_i} \ , \cr
Q_{\alpha_{2i-1}} \ &= \ \oint dz~
(\psi_i ~-~ \psi_{i-1})~e^{{i  \sqrt{k+n+1}} ~ (\phi_i - \phi_{i-1})}
\ , } }
with the convention that $\phi_0 \equiv \phi_{n+1} \equiv 0$,
$\psi_0 \equiv \psi_{n+1} \equiv 0$.

The higher spin $W$-generators can, in principle, be extracted from
\laxope\ and rewritten in terms of the $N=2$ superfields.  In
practice,
this can be algebraically very cumbersome, and has only been done for
the model \coset\ with $n=2$.  In this model, the spin-$2$ superfield
may be written explicitly as:
\eqn\Wdefn{ \eqalign{{\cal W} \  &= \ \widetilde
W(z,\theta^+, \theta^-) ~-~ \coeff{1}{4}~\alpha_0^2 \del
{\cal J}(z,\theta^+,\theta_-) ~-~ {3-8\alpha_0^2\over
2(5-18\alpha_0^2) } ~ :{\cal J}^2(z,\theta^+,\theta^- ): \cr
&  \qquad ~+~  { (1 - 3\alpha_0^2 )(1 + 2 \alpha_0^2) \over
8(5 - 18 \alpha_0^2) }~(D^+D^--D^-D^+)~
{\cal J}(z,\theta^+,\theta^-) \ ,\cr } }
where
\eqn\feltildeww{\eqalign{
&\widetilde W(z,\theta^+,\theta^-) \ =  \
- i {\alpha_0^3 \over 4}\del^2\Phi_1^+
+ i { \alpha_0^3 \over 4} \del^2\Phi_1^- + i { \alpha_0^3 \over 4 }
\del^2\Phi_2^- + {\alpha_0^2 \over 8} D^+\del \Phi_1^+ D^-\Phi_1^-
\cr
& \ \ \ + { \alpha_0^2\over 8 }D^+\Phi_1^+\del D^-\Phi_1^-  +
{\alpha_0^2 \over 8 } D^+\del\Phi_1^+D^-\Phi_2^-
+ { \alpha_0^2\over 4 }  \del\Phi_2^+ \del\Phi_1^-
 + { \alpha_0^2\over 4 } \del \Phi_2^+ \del\Phi_2^- \cr
 & \ \ \ - {\alpha_0^2
\over 4 }
\del\Phi_2^- \del\Phi_1^-    - { \alpha_0^2 \over 4
}\del\Phi_2^-\del\Phi_2^- -
{ \alpha_0^2 \over 4 } \del\Phi_2^+\del\Phi_1^+ + { \alpha_0^2 \over
4 }
\del\Phi_2^- \del\Phi_1^+
 \cr
 & \ \ \  - i { \alpha_0 \over 8 }\del\Phi_2^+D^+\Phi_1^+D^-\Phi_1^-+
i { \alpha_0 \over 8 } \del\Phi_2^- D^+\Phi^+_1
D^-\Phi_1^-
 + i { \alpha_0\over 8 } \del \Phi_1^- D^+\Phi_2^+ D^-\Phi_2^-  \cr
 & \ \ \  +
i {\alpha_0 \over 8 } \del \Phi_2^- D^+\Phi_2^+ D^- \Phi_2^- + i {
\alpha_0 \over 8 }\del\Phi_1^- D^-\Phi_2^-
D^+\Phi_1^+
-i{\alpha_0 \over 8 } \del\Phi_1^+ D^+\Phi_2^+D^-\Phi_2^- \cr
& \ \ \ +
{ 1 \over 16 } D^+\Phi_1 D^-\Phi^-_1D^+\Phi^+_2 D^-\Phi_2^-  \ . \cr
}}
If we define
\eqn\defhatjj{\eqalign{\hat {\cal J} \ = \ & +
\coeff{1}{\sqrt{2(k+2)}}~
(\del \Phi^+_1-\del \Phi^+_2-\del \Phi^-_1) \cr &
 +~ \coeff{i}{ 2 \sqrt 2} \sqrt{\coeff{k+3}{k+2}}~
(D^+\Phi^+_1D^-\Phi^-_1-  D^+\Phi^+_2D^-\Phi^-_2)}}
Then we can write \feltildeww\ as
\eqn\callw{\eqalign{ \widetilde{\cal W}\ = \ &  \coeff{1}{4}
{\cal J}^2 ~+~ \coeff{1}{8(1-\alpha^2_0)} ~\hat {\cal J}^2 ~+~
\coeff{ \alpha^2_0}{ 4}\Bigl ( \del { \cal J} ~+~ i
\coeff{1}{2(1-\alpha_0^2)}~
\hat\del {\cal J} \Bigr ) ~+ \cr & \coeff{\alpha^2_0}{8  }~
D^+\del \Phi^+_1D^-\Phi^-_2 ~+~ \coeff{i \alpha_0}{ 8 }~
\del \Phi^-_1 D^-\Phi^-_2D^+\Phi^+_1 \  .} }
Combining eq. \Wdefn\ and \callw\ we have
\eqn\neweq{\eqalign{ {\cal W} \ = \ & \coeff{c_2}{2} (D^+D^--D^-D^+)
{\cal J} ~-~ \coeff{6c_2}{ c } ~{\cal J }^2\cr
 & -~ (c_1+c_2)\Bigl( -
\coeff{1}{2}~ \hat {\cal J}^2 {}~+~ \coeff{ i\alpha_0^2}{
\sqrt{2(1-\alpha_0^2)} } ~\del \hat {\cal J} \Bigr)\cr & +~
\coeff{\alpha_0^2}{2 \sqrt {2(1-\alpha_0^2)}} ~D^+\del
\Phi^+_1D^-\Phi_2^- \cr
&+~ \coeff{i \alpha_0^2}{ 2(1-\alpha_0^2)}~
\del \Phi^-_1D^-\Phi^-_2D^+\Phi_2^+, }}
where
\eqn\defcc{c_2\ = \  {(1-3\alpha_0^2)(1+2\alpha_0^2)\over
(1-\alpha_0^2)}
\  , \ \ \
c_1+c_2 \ = \  5-18\alpha^2_0  \ \  \ \ {\rm and} \ \  \ \
c \ = \ 6(1-3\alpha^2_0 )}

\subsec{The \LG free field formulation}

\nref\LGboth{E.~Martinec, \plt{217B} (1989) 431;
C.~Vafa and N.P.~Warner, \plt{218B} (1989) 51.}
\nref\Morozov{A.~Marshakov  and A.~Morozov, ``Landau-Ginzburg models
with $N=2$ supersymmetry as conformal theories,'' ITEP preprint,
ITEF-89-183 1989.}
\nref\HLPM{H.C.~Liao and P. ~Mansfield \plt{B255} (1991) 237.}
\nref\MGDZ{M.T.~ Grisaru  and D.~Zanon, \plt{B332} (1994) 77.}

The idea in this formalism is to directly use the $N=2$
supersymmetric Landau-Ginzburg model with action:
\eqn\susyact{ S \ = \ \int \ d^2  x \ d^4 \theta \ \sum_j  {
\Phi^+}_j ~ \Phi^-_j  ~-~ \int \  d^2 x \ d^2 \theta\  W(
\Phi_j^+) \ ~-~ \ \int \ d^2 x \ d^2 \bar \theta \ W( {
\Phi}^-_j)
\ , }
where $ \Phi^\pm_j$, $j=1,\ldots , n$ are $N=2$ (anti)-chiral
superfields.
If $W$ is quasihomogenous then the \LG model
\susyact\ with its ``trivial'' kinetic term is superconformally
invariant on the cohomology of the half of the supercharges \WitLGb.
This is in the same spirit as the work of \refs{\LGboth {--} \MGDZ}
in that there is certainly a kinetic term that renders the model
exactly superconformal, and such a kinetic term can be viewed as a
cohomologically trivial correction to that of \susyact. It
was also shown in \WitLGb\ that the superconformal generators could
be identified using the equations of motion of \susyact\ alone.
Indeed \susyact\ implies that the fields $ \Phi^+_j$ and ${
\Phi^-}_j$ have logarithmic short distance expansion, and the
left-moving $N=2$ superconformal stress energy tensor can be
represented by:

\eqn\freej{ {\cal J} \ = \ \sum_j  \Big[\ \coeff{1}{4} (1-\omega_j)
D^+  \Phi^+_j  D^-{ \Phi}^-_j -
 \ \coeff{1}{2}\omega_j  \Phi^+_j \del  { \Phi}^-_j
\ \Big] \ \ , }
where the $\omega_j$ are the scaling dimensions of the \LG fields
$ \Phi^+_j$.  For the model \coset\ they are given by
$\omega_j = {j \over k+n+1}$.
The current ${\cal J}$ has been constructed so as to satisfy
\eqn\currentc{\bar D^-{\cal J} \ = \ 0 \ \ , }
given the equations of motion of \susyact.

This is closely related to the free field approach of
\ref\FGLS{P.\ Fr\'e, L.\ Girardello, A.\ Lerda and P.\ Soriani,
\nup{387} (1992) 333.}.  That is, one can describe the \LG system
in terms of twisted ghost and superghost fields.  Introduce
anti-commuting fields $\hat b_j(z)$ and $\hat c_j(z)$, and
commuting fields $\hat \beta_j(z)$ and $\hat \gamma_j(z)$,
with operator products:
\eqn\ghopes{\hat b_i(z) ~ \hat c_j(w) ~\sim~ {\delta_{ij}\over z-w}
\qquad \hat \beta_i(z) ~ \hat \gamma_j(w) \ \sim \ -
{\delta_{ij}\over z-w} \ .}
The superconformal generators are then:
\eqn\ghostsca{\eqalign{J(z)\ = \ &  - ~\sum_{j=1}^n \big[
(1 -  \omega_j) \hat b_j ~ \hat c_j ~-~ \omega_j \hat \beta_j
\hat \gamma_j \big ]  \cr
G^{+}(z) \ = \ &  ~\sum_{j=1}^n \big[  (1 -  \omega_j) \hat c_j~
\del \hat \beta_j~-~\omega_j \hat \beta_j ~ \del \hat c_j \big ]\ ;
\quad G^{-}(z) \ = \  ~\sum_{j=1}^n ~ \hat b_j \hat \gamma_j
\cr
T(z) \ = \ & - \shalf \sum_{j=1}^n \big[(1 +  \omega_j)\hat b_j ~
\del
\hat c_j ~+~ (1 - \omega_j) ~\hat c_j ~ \del \hat b_j  ~+~  \omega_j
\hat \beta_j \del \hat \gamma_j \cr
& -~  (2  - \omega_j ) \hat \gamma_j
\del
\hat \beta_j \big ] \ . \cr }}
The fields $\hat \beta_j(z)$ and $\hat b_j(z)$ can be identified
with the bosonic and and fermionic components of the superfield
$\Phi_j^+$, while $\hat \gamma_j$ and $\hat c_j(z)$ can be
identified with the components of $\Phi^-_j$.

One can easily determine the relationship between the the \LG
fields and the free fields of the last subsection by using the
dimensions and charges of the \LG fields along with the fact that
\ghostsca\ must be the same as \ntwogen.  From this we find
\eqn\transl{\eqalign{\hat \beta_j &  \ =
\ e^{i \alpha_0  \phi_j } \ ;
\qquad \hat \gamma_j \ = \ ( \psi_j \bar \psi_j ~+~ i \sqrt{k+n+1}~
\del  \bar \phi_j) ~e^{- i \alpha_0  \phi_j } \cr
\hat b_j & \ = \ -~\coeff{i}{\sqrt{k+n+1}}~ \psi_j~  e^{ i \alpha_0
\phi_j} \ ; \qquad \hat c_j \ = \ - i \sqrt{k+n+1}~\bar \psi_j~
e^{ - i \alpha_0 \phi_j} \ . \cr }}

The shortcoming of the \LG motivated free field formulation is
that the \LG formulation provides one with
very little information about the screening currents.  From
\refs{\FGLS,\WitLGb,\DNNWgen} it is evident that such knowledge
is unnecessary if one wants to study the topological matter model
or extract the elliptic genus.  However, the screeners are essential
in order to get the complete conformal theory.  Using \transl\ one
could, at least in principle, obtain the proper \LG screeners
from the screeners of the Drinfeld-Sokolov reduction.

It is also interesting to observe that in the complete \LG theory,
the holomorphic supercurrent $G^+(z)$, and its anti-holomorphic
counterpart, $\bar G^+(\bar z)$, receive corrections from the
superpotential. Indeed the complete supercurrent with
anti-holomorphic component $\bar G^+(\bar z)$ has a holomorphic
component that can be written:
\eqn\potscreen{\sum_{j=1}^n ~{ \del W(\Phi_\ell^+)\over\del
{\Phi_j^+ }}  ~ D_- \Phi^+_j ~\bigg|_{\theta =\bar
\theta =0} \ . }
These currents appear to be the \LG analogue of the
F-type screening currents in the Drinfeld-Sokolov reduction.
In the $N=2$ superconformal minimal model (with one
superfield) the identification is exact, but the
precise relationship is rather less clear for the more
general models.

One can now use \transl\ to translate the $W$-algebra generators of
the previous section into the \LG formulation. Alternatively, one can
obtain these $W$-generators by making an Ansatz, imposing chirality
of the $W$-superfield and using the \LG equations of motion as in
\refs{\KMohri,\DNNWgen}. In the appendix to this paper we give
details of such a computation for the first $W$-superfield for the
\LG theory with two fields. This computation, along with the
foregoing translation to the Drinfel'd-Sokolov formulation, lead us
to believe that the process of imposing chirality and the operator
equations of motion in the \LG formulation is basically equivalent to
imposing commutation with the fermionic screening charges in the
Drinfel'd--Sokolov reduction, and so the chirality and the \LG
equations of motion are, in principle, a little less stringent than
the requirements of the full Coulomb gas description. In practice,
for the Ans\"atze that we have used, chirality and the \LG equations
of motion are sufficient to determine the $W$-generator. However, the
process of constructing the quantum versions of the $W$-generators
using the \LG formulation is operationally more difficult to
implement, and it is easier to use the Drinfel'd--Sokolov reduction
(along with the simplifications to be discussed in the next section).

\subsec{Coulomb Gas formulations of related coset models}

The Coulomb gas formulation that we will discuss here can only be
properly justified by the results of the next section, and we include
it here for completeness. The idea is to find free bosonic
descriptions of the factors in the tensor product \tensor.

\nref\BMP{P.~Bouwknegt, J.~McCarthy and K.~Pilch,\nup{352} (1991)
139.}

There is a well known, standard Coulomb gas description of
${\cal M}_2$ in terms of free bosons \refs{\BilGe {--} \BMP}.
Let $\sigma(z)$ denote a vector of $n-1$ canonically normalized
free bosons with energy-momentum tensor:
\eqn\Ttwo{T_{2}(z) ~=~ - \half (\partial \sigma(z) )^2 ~+~
i ~(\beta_+ ~-~ \beta_-) ~\rho \cdot \partial ^2 \sigma(z) \ ,}
where $\rho$ is the Weyl vector of $SU(n)$ and
\eqn\betadef{\beta_\pm ~\equiv~ ~ \Bigg[
\sqrt{{{k + n + 1} \over {k + n}}}~ \Bigg]^{\pm 1}\ .}
The screening currents are then
\eqn\todascreen{S^\pm_{\gamma_j} ~=~ e^{\pm i \beta_\pm \gamma_j
\cdot \sigma (z)} \ ,}
where the $\gamma_j$ are the simple roots of $SU(n)$.

The highest weight fields of $ {\cal M}_2$ can be represented as:
\eqn\highwei{ V_{\lambda_+, \lambda_-} ( z) \ = \  e^{ - i(\beta_+
\lambda_+ - \beta_- \lambda_-) \cdot \sigma(z) } \  . }
This has conformal weight
\eqn\conwei{\eqalign{\Delta_{\lambda_+, \lambda_-} \ = \ & \half
(\beta_+
\lambda_+ - \beta_-\lambda_-)^2 ~+~  (\beta_+ - \beta_-) \rho \cdot
(\beta_+ \lambda_+ ~-~ \beta_- \lambda_-) \cr
\ = \ &  { \lambda_+ \cdot (\lambda_+ + 2\rho) \over 2(k+n) } ~+~
\half ( \lambda_+ - \lambda_-)^2 ~-~ { \lambda_- \cdot (\lambda_-
+ 2\rho) \over 2(k+n+1) } \ . \cr }}
Thus $\lambda_+$ and $\lambda_-$ can be thought of as corresponding
to the weights of the $SU_k(n)$ and $SU_{k+1}(n)$ factors of
${\cal M}_2$.

The model ${\cal M}_1$ can be realized in terms of $2n$ free bosons.
Let $\chi$ and $\xi$ be vectors of $n$ canonically normalized free
bosons,
and take
\eqn\Tone{T_1(z) \ = \ -\half (\del \xi)^2 ~-~ \half (\del\chi)^2
{}~-~ {i \over \sqrt {k+n+1}}~\tilde \rho \cdot \del^2 \xi ~-~
{1 \over \sqrt {k+n}} ~\rho \cdot \del^2 \chi   \ \ ,  }
where $\tilde \rho$ is the Weyl vector of $SU(n+1)$.
The natural choice for representations of the highest weight fields
are
\eqn\repfield{U_{\lambda_+,\lambda_-} \ = \ exp \Big[ { + i ~
{\lambda_+ \cdot \xi \over \sqrt {k+n+1}} ~+~ {\lambda_- \cdot
 \chi \over \sqrt {k+n}}
 \Big]   }  \ . }
This has conformal weight
\eqn\connweight{{\lambda_+ \cdot (\lambda_+ +2\tilde \rho ) \over
2(k+n+1) } -
{\lambda_- \cdot (\lambda_- +2 \rho ) \over 2(k+n) }    \ \ ,}
which is consistent with with identifying $ \lambda_+$ and $
\lambda_-$
with heighest weights of the numerator and denominator factors
respectively of
${\cal M}_1$.
The screening  currents are somewhat more difficult to determine,
and will be given in the next section.

\newsec{The $N=2$ super-$W$ structure and factorizing the Coulomb gas
description}

It was observed in section 3.1 that the simplest way to get at the
generators of the $N=2$ super-$W$ algebra is to find the
supercharges,
and the $W$-generators of the model ${\cal M}_2$ in \tensor.  We will
therefore show explicitly how the Coulomb gas descriptions of the
last section decompose into a tensor product.  We will also have to
handle the subtleties described in section 2.

The key to extracting the bosonic formulations of the factor models
in \tensor\ from the Drinfeld-Sokolov reduction is to use the
screening charges.
Modulo the subtleties of section 2, the screening charges \anscren\
and \screanone\ must be sums of screening charges for the factor
models.
(We will discuss the role of the fermionic screeners, \screenann,
later.)
Moreover, the roots of the $A_n$ and $A_{n-1}$ subalgebras of the
superalgebra $A(n,n-1)$ should coincide with the roots of the factors
of $A_n$ and $A_{n-1}$ in \tensor.  This leads to the following
fairly unambiguous identification:
\eqn\sutopid{\gamma_i \cdot \xi ~\equiv~ (\alpha_{2i-1} +
\alpha_{2i}) \cdot \varphi ~\equiv~ \phi_i + \bar \phi_i -
\phi_{i-1}\ , \quad i = 1, \dots,n \ ;}
where the $\gamma_i$ are the simple roots of $SU(n+1)$, the
$\alpha_j$
are the simple roots of $A(n,n-1)$, and the bosons $\xi$ are those of
section 3.4.

The simple roots of an $A_{n-1}$ subalgebra are given by
$\alpha_{2i} + \alpha_{2i+1}$, and one would expect this to coincide
with a linear combination of the bosons, $\sigma$, of ${\cal M}_2$,
and the bosons $\chi$ of the denominator of ${\cal M}_1$.  To isolate
bosons corresponding ${\cal M}_2$ one seeks the screening currents
corresponding to the denominator of ${\cal M}_2$, that is, those
screening currents that have monodromy involving $(k+n+1)^{\rm th}$
roots of unity.  From \modscreen, and its obvious generalizations,
one sees that the screeners that can be modified (as in section 2)
are those associated
with the locked numerator factor of ${\cal M}_2$, while the other
screeners ar unchanged.  Noting that the screening charges in
\newscreener\ involves the $(k+n+1)^{\rm th}$ roots of unity, it
is natural to look for the pure vertex operator screeners,
$S^-_{\gamma_j}$, of \todascreen\
in the $SU(n)$ screener, $Q_{\alpha_{2i} + \alpha_{2i+1}}$, of
\newscreener.  From this it is not hard to identify the second
fermion bilinear term as the one we want.

Bosonize the fermions according to:
\eqn\bosonize{\eqalign{& \psi_j(z) ~=~ e^{i H_j(z)} \ , \qquad
\bar \psi_j(z) ~=~  - e^{- i H_j(z)} \ , \cr
& \bar \psi_j(z) ~ \psi_j(z) ~=~ i \del H_j(z) \ , \qquad
\psi_j(z) ~ \bar \psi_\ell(z) ~=~  e^{i( H_j(z) - H_\ell(z))}
\ , \ j \neq \ell \ ;}}
where  $H_j(z) ~H_\ell(w) \sim - \delta_{j \ell}~ log (z-w)$.
Writing the second fermion bilinear of
$Q_{\alpha_{2i} + \alpha_{2i+1}}$ as a pure vertex operator,
we can then identify the free bosons of ${\cal M}_2$:
\eqn\minid{\gamma_j \cdot \sigma ~ \equiv ~
 \sqrt{\coeff{k+n+1}{k+n}} ~(H_j - H_{j+1})~+~\coeff{1}{\sqrt{k+n}}~
(\phi_j - \bar \phi_j  - \phi_{j+1}) \ , \quad j = 1, \dots, n-1 \ .}
The last $U(1)$ factor in \tensor\ is the $N=2$, $U(1)$ current,
which can be written:
\eqn\uonecur{J(z) ~=~ i~ \sum_{j=1}^n \big[~ \del H_j(z) ~+~
\coeff{1}{\sqrt{k+n+1}}~ (~ \del \phi_j ~-~ j ~\del \bar \phi_j ~)
\big ] \ .}
The remaining bosons of ${\cal M}_1$ are the natural orthogonal
combinations to \sutopid, \minid\ and \uonecur.  This yields the
identifications:
\eqn\chiid{\eqalign{\gamma_j \cdot \chi & ~\equiv~
\coeff{i}{\sqrt{k+n}} ~ (H_j - H_{j+1}) ~+~ i~
\sqrt{\coeff{k+n+1}{k+n}} ~(\phi_j -  \bar \phi_j  - \phi_{j+1})
\ , \quad j = 1, \dots, n-1 \cr
K(z) & ~\equiv~ 2i~\sqrt{\coeff{k}{n(n+1)}}~ (\tilde \rho - \rho)
\cdot \del \chi \cr
&~\equiv~ \sqrt{\coeff{n+1}{n}} ~\sum_{j=1}^n
\Big[~ \del H_j(z) ~+~ \coeff{\sqrt{k+n+1}}{n+1}
{}~(~ \del \phi_j ~-~  j~\del \bar \phi_j ~) \Big ] \ .} }
The current, $K(z)$, corresponds to the $U(1)$ factor in
${\cal M}_1$ and has been normalized according to
$K(z)~K(w) \sim {k \over (z-w)^2}$ .

One can now rewrite the entire model in terms of these free
bosonic fields.  The $W$-generators of ${\cal M}_2$ can be written
in the usual manner as Weyl invariant combinations of the
derivatives of the bosons $\sigma$ \refs{\EY{--}\FatL}.  This enables
one to write down rather explicit expressions for the
bottom components of the super-multiplets.  There are, however,
the subtleties discussed in section 2.

One can easily express the D-type screening currents in the new
free bosonic basis.  To do this, it is convenient
to introduce the standard basis, $e_j$, $j =1, \dots, n+1$ for the
weight
space of $SU(n+1)$.  The vectors $e_j$ satisfy:  $\gamma_j =e_j -
e_{j+1}$, $\sum_{j=1}^{n+1} e_j =0$ and
$e_i \cdot e_j = \delta_{ij} - { 1 \over (n+1)}$.
Introduce a similar basis $\hat e_j$ for $SU(n)$.  The
vectors $\hat e_j$ are orthogonal to $e_{n+1}$ and are given by
$\hat e_j = e_j - {1 \over n} e_{n+1}$, $j = 1, \dots, n$.
Using these vectors the screening currents corresponding to
$Q_{\alpha_{2j-1} + \alpha_{2j}}$ and
$Q_{\alpha_{2j} + \alpha_{2j+1}}$
can be respectively written as:
\eqn\transscreena{\eqalign{ U_j(z) ~&=~ \coeff{2i}{\sqrt{k+n+1}}
{}~ exp{\Big[-\coeff{i}{\sqrt{k+n+1}} ~(\gamma_j \cdot \xi) -
\coeff{1}{\sqrt{k+n}} ~(\gamma_{j-1} \cdot \chi) \Big] } \cdot\cr
& exp{\Big[+i \sqrt{\coeff{k+n+1}{k+n}} ~(\gamma_j \cdot \sigma)
 ~\Big] }
 \quad  ~-~ i \sqrt{k+n+1}~ \del ~\Big(~
e^{- \coeff{i}{\sqrt{k+n+1}} ~\gamma_j
\cdot \xi} ~ \Big) \cr
& -~ 2 ~\Big[ ~e_{j+1} \cdot \del \xi(z)
  \qquad ~-~ i~\sqrt{\coeff{k+n}{k+n+1}}~
\big(~\hat e_j \cdot  \del \chi(z) ~\big) \cr
& +~  \coeff{1}{\sqrt{(k+n+1)n(n+1)}} ~K(z)   ~\Big]\cdot
 e^{- \coeff{i}{\sqrt{k+n+1}} ~\gamma_j \cdot \xi} \ ,
\quad j ~=~ 1, \dots, n \ ; \cr}}
\eqn\transscreenb{\eqalign{
V_j(z) ~&=~ \coeff{2i}{\sqrt{k+n+1}}~ exp{\Big[~-i~
\sqrt{\coeff{k+n}{k+n+1}} ~(\gamma_j \cdot \sigma) ~\Big] } \cr
{}~&+~ \quad i ~\sqrt{k+n+1} ~ \del ~ \Big(~
exp \Big[~-~ \coeff{1}{\sqrt{k+n}} ~(\gamma_{j} \cdot \chi) \cr
& +~ \coeff{i}{\sqrt{(k+n+1)(k+n)}} ~(\gamma_j \cdot \sigma)
{}~\Big]  ~\Big) \cr
{}~&-~ 2 ~\Big[ ~e_{j+1} \cdot \del \xi(z)  ~-~
i~\sqrt{\coeff{k+n}{k+n+1}}~ \big(~\hat e_j \cdot \del
\chi(z) ~\big)  \cr
&   +~ \coeff{1}{\sqrt{(k+n+1)n(n+1)}} ~K(z) ~ \Big]~
exp \Big[- \coeff{1}{\sqrt{k+n}} ~(\gamma_{j} \cdot \chi) \cr
& + \coeff{i}{\sqrt{(k+n+1)(k+n)}} ~(\gamma_j \cdot \sigma)
{}~\Big] \  ,
 \quad j ~=~ 1, \dots, n-1 \ \cr .}}

These screening currents have precisely the kind of structure
that was described in section 2.  That is, the screening
currents of one factor of the tensor product \tensor\
have been mixed
with dual representatives of the vacuum of the other factor in
the tensor product.  The operators analogous to \Xdefn,
that extend the chiral algebra of the bosonized theory, are
nothing other than combinations of derivatives of $\phi_\ell$
and $\bar \phi_\ell$ with nilpotent fermion bilinears like
$\psi_i \bar \psi_{j}$ and $(\del \psi_i) \bar \psi_{j}$.
We should therefore expect such
corrections to the standard forms of the $W$-generators.
Indeed, the corrections to $S(z)$ analogous to $R^+S^-$ in
\repS\ can be read off from \transscreena\ and \transscreenb.
These terms are of the form:
\eqn\Nilpj{\eqalign{N_j(z) ~&\equiv~ \Big[ ~e_{j+1} \cdot
\del \xi(z) ~-~ i~\sqrt{\coeff{k+n}{k+n+1}}~
\big(~\hat e_j \cdot \del \chi(z) ~\big) ~-~
\coeff{1}{\sqrt{(k+n+1)n(n+1)}} ~K(z) ~ \Big] \cr
& \qquad \qquad \qquad \times~ exp \Big[~-~ \coeff{1}{\sqrt{k+n}} ~
(\gamma_{j}  \cdot \chi) ~+~ i\sqrt{\coeff{k+n+1}{k+n}}~
(\gamma_j \cdot \sigma)~ \Big] \cr
&\equiv~ \big[~ (\del \bar \phi_j)~ \psi_j ~\bar \psi_{j+1} ~+~
i \alpha_0 ~(\del \psi_j) ~\bar  \psi_{j+1} ~\big]  \ . }}
These terms must be added (with appropriate coefficients)
to the naive form for $S(z)$, and for $n \geq 3$ there will be
further terms of the form
\eqn\moreNilp{\big[~ (\del \bar \phi_j)~ \psi_j ~\bar \psi_{j+\ell}
{}~+~ i \alpha_0 ~(\del \psi_j) ~\bar  \psi_{j+\ell} ~\big]  \ .}
These terms are necessary to cancel other terms that result from
commuting the screening charges with the $N_j(z)$.  Alternatively,
they have to be present for $S(z)$ to have the proper operator
product with itself.

Rather than get too deeply involved in the technical details of the
general problem, we will specialize to the model \coset\ with
$n=2$.  The model, ${\cal M}_2$, is then an ordinary minimal
($c < 1$) model, and is realized by a single free boson $\sigma$.
The energy momentum tensor of the complete $N=2$ supersymmetric
model is, of course,
the simple sum of all the component energy momentum tensors:
\eqn\sumT{T_1(z) ~+~ T_2(z) ~+~ \coeff{k+3}{12k} ~ J^2(z) \ ,}
where $T_1$ and $T_2$ are given by \Tone\ and \Ttwo. The bottom
component, $S(z)$, of the $W_3$-supermultiplet is a spin-$2$ current,
and the naive guess for its form is \simpleS. As explained
earlier, even though $T_2$ is not a good conformal field, it can be
viewed as defining the extension of the chiral algebra, and once one
has it, one can easily construct $S(z)$.  The proper representative
of $T_2$ in the tensor product model will involve a correction of
the form \Nilpj.  Indeed, we find that the complete free-field
expression for $T_2(z)$ is:
\eqn\corrTtwo{\eqalign{\widehat T_2(z) ~=~ - \half (\partial
\sigma(z) )^2  &~+~ \coeff{i}{\sqrt{2(k+2)(k+3)}} ~ \partial^2
\sigma(z)  \cr
&~+~ \coeff{k+3}{2(k+2)} \big[~ (\del \bar \phi_1)~ \psi_1
{}~\bar \psi_{2}  ~+~ \coeff{i}{\sqrt{k+3}}  ~(\del \psi_1) ~
\bar  \psi_{2} ~\big] \ .}}
The coefficient of the fermion bilinear terms is determined by
requiring that $\widehat T_2(z)$ commute with the D-type screeners.
The extra nilpotent fermion bilinears are, of course, present in
\feltildeww: the two relevant terms are the bottom components of
${\alpha_0^2 \over 8 } D^+\del\Phi_1^+D^-\Phi_2^-$ and
$i {  \alpha_0 \over 8 }\del\Phi_1^- D^-\Phi_2^- D^+\Phi_1^+$.

To summarize, the key to extracting the $W$-algebra generators
is in the identification of the free bosons, $\sigma$, given in
\minid.  The lowest component of each $W$-supermultiplet can
then easily be constructed from them.  Since we will need it
later, we conclude by giving the form of these bosons in terms of
the \LG free fields.  Indeed, from \minid\ and \transl\ one easily
obtains:
\eqn\ghminid{\gamma_j \cdot \sigma ~=~ \sqrt{\coeff{k+n+1}{k+n}}~
\Big[ \big(1 - \coeff{1}{k+n+1} \big)~ \hat b_j \hat c_j ~-~
\hat b_{j+1} \hat c_{j+1} \Big] ~-~ \coeff{1}{\sqrt{(k+n+1)(k+n)}} ~
\hat \beta_j \hat \gamma_j \ .}

\newsec{The elliptic genus and other characters}

One of the beautiful features of the \LG model is that the
elliptic genus of the model can be easily computed solely from
the knowledge of the field content and scaling dimensions
\WitLGb.   This computation can be refined so as to
determine the the $U(1)$ eigenvalues of the states contributing
to the elliptic index \refs{\PDFSY{--}\MHenn}.  One of the basic
ideas of \refs{\KMohri,\DNNWgen,\WLAS} was that the \LG potential
contains the information about when the conformal model has an
extended chiral algebra.  This fact was further employed in
\DNNWgen\ to show how, at least for the model \coset\ with $n=2$,
the elliptic genus could be further refined so as to extract exactly
how the different eigenstates of the extended chiral algebra
contribute to the elliptic index of the model. The result, for
\coset\ with general $n$, was also conjectured in \DNNWgen, and in
the last section we have developed enough information to now show
that this conjecture is correct.

\subsec{The refined elliptic genus}

The idea is to introduce the function:
\eqn\ellchar{{\cal F} (q,\mu,\nu) ~=~ Tr_{\cal H} ~\left( (-1)^F
q^{H_L}~ \bar q^{H_R} ~ exp(i \mu \cdot j_{0}) ~ exp(i \nu J_{0})
\right)  \ .}
In this expression $\cal H$ is the complete Hilbert space of the
model in
the Ramond sector, $H_L = L_0$ and $H_R = \bar L_0$ are the
hamiltonians
of the left-movers and right-movers, $F$ is the total fermion number,
$J_0$ is the left-moving $N=2$, $U(1)$ charge, and $j_0$ is the
vector
of zero modes of the left-moving bosons, $\sigma$, defined in
\ghminid.
The standard index argument can be used to show that in the
right-moving
sector, only the ground-states contribute to the trace.  As a result,
the function ${\cal F}$ is a function of $q$ alone (and not a
function
of $\bar q$), and consists of a sum of the (left-moving) Ramond
ground-state  characters.  Unless one sets $\mu \equiv 0$, the result
will {\it not} be characters of \coset.  This is because the
charges $j_0$ do not commute with the screening charges that reduce
the
free field Hilbert space down to that of the coset model.

However, it was argued in \DNNWgen\ that one can obtain a
character of the coset model by the simple expedient of symmetrizing
with respect to the Weyl group of $SU(n)$.  That
is, one defines
\eqn\symchar{{\cal F}_s (q,\mu,\nu) ~=~ \sum_{w \in W(SU(n))} ~
 {\cal F} (q,w(\mu),\nu)  \ .}
To see why this is so, one first evades all the subtleties of
section 2 by simply deciding to describe the model as a naive
tensor product \tensor, and not as a locked tensor product with
non-standard screening charges.  That is, one uses the same set of
free fields, but simply chooses the naive set of screening
charges for a tensor product model.  The cost of doing this is that
one must remember to lock the Hilbert spaces together by hand once
one has constructed them from the free fields.  The advantage of
taking the naive tensor product is, of course, that the
$W$-generators
have the simple polynomial form in derivatives of the free bosons.

If one temporarily ignores the oscillator contributions to the
bosonic Hilbert space of ${\cal M}_2$ of \tensor, one can see,
by performing integral transforms as in \DNNWgen, that ${\cal F}_s$
contains the same information as refining the elliptic genus with
respect to the zero-modes of the $W$-algebra of ${\cal M}_2$.
This is simply a version of the theorem that a weight of a
Lie algebra is uniquely specified, {\it  up to Weyl rotations},
by the values of all the Casimirs on that weight.
The problem is with the oscillator contributions.  The zero-modes
of $W$-generators are notorious for only really being diagonalizable
on pure momentum states.   We do not know how
to make a compelling argument solely from the perspective of the
\LG formulation.  However,  based on the results of the last section,
we know that the bosons $\sigma$ are precisely those of the standard
Coulomb gas formulation of ${\cal M}_2$.  These characters consist of
trivial oscillator $\eta$-function factors multiplying sums over
pure $\sigma$-momentum states.   Thus the null states introduced
by the full screening charges of \tensor\ only involve the pure
momentum
states and are thus correctly reproduced in \symchar.
As a result, the Weyl symmetrized ${\cal F}$ will suffice to produce
a function on the Hilbert space of ${\cal M}_2$ and hence on the
Hilbert
space of \coset.

The argument can be made rather more directly if one merely
concentrates upon the zero mode, $S_0$, of $S(z)$. Since this is a
linear combination of the energy momentum tensors in the tensor
product, this grades the elliptic genus according to the energies
associated with the factors in \tensor. (It is the higher spin
generators of the $W$-algebra that cause the problems with
simultaneous diagonalization.) As was observed in \DNNWgen, this
refinement of the elliptic genus is related fo ${\cal F}_s$ by a
Laplace transform. The oscillator parts are dealt with by multiplying
by ratios of $\eta$-functions in such a manner as to reflect the fact
the $n-1$ of the bosons have their energies measured in ${\cal M}_2$.
Thus the component parts of the tensor product can be factored out
easily from ${\cal F}_s$.

To be more specific, the ${\cal M}_2$ components
of \symchar\ will consist of functions of the form:
\eqn\Mtwochar{\eqalign{\chi_{\lambda_+}^{\lambda_-}\ = \ & {1\over
\eta(\tau)^\ell} \sum_{w\in W(SU(n))} \sum_{ \gamma \in M(G)}
\epsilon(w)\cr
& \times  q^{\half \big [ \beta_+   w(\lambda_+
+ \rho) ~-~ \beta_- (\lambda_- + \rho) ~+~
\sqrt{(k+n)(k+n+1) } ~ \gamma  \big ]^2} \cr
&\times e^{i \mu \cdot \big [ \beta_+   w(\lambda_+
+ \rho) ~-~ \beta_- (\lambda_- + \rho) ~+~
\sqrt{(k+n)(k+n+1) } \gamma  \big ]} \ . \cr }}
For $\mu \equiv 0$ these functions are characters of the model
${\cal M}_2$.  Therefore, if we extract the coefficient
of
\eqn\phaseform{
e^{i \mu \cdot \big [ \beta_+   w(\lambda_+ + \rho) ~-~ \beta_-
(\lambda_- + \rho) ~+~ \sqrt{(k+n)(k+n+1) }   \gamma  \big ]} ~
\times~ e^{i \nu a} \ ,}
in ${\cal F}_s$, then we will obtain the character of the model
${\cal M}_1$ that is paired with the states in the Hilbert space of
${\cal M}_2$ labelled by $\chi_{\lambda_+}^{\lambda_-}$, and which
also have $N=2$, $U(1)$ charge equal to $a$. While we have not
rigorously proved the foregoing statement, we think it is emminently
plausible, and in the next sub-section we will confirm our results by
computing branching functions in the factors of \tensor.

Thus the refined elliptic genus, ${\cal F}_s$, enables  us to
completely decompose and isolate the component parts of the
partition function of \coset.

\subsec{Explicit formulae and a simple example}

Following the arguments of Witten, we know that
the elliptic genus can be expressed very simply in terms of the
free fields in the Ramond sector.  That is,
it can be expressed as a simple product
of ratios of theta functions.  The refined ``elliptic character,''
${\cal F}$, is obtained by grading this product of theta functions
with the bosonic zero-modes $J_0$ and $j_0$.  Using \ghostsca\ and
\ghminid\ we obtain the following formula for ${\cal F}$:
\eqn\Fform{{\cal F} (\tau, \mu, \nu) ~=~ \prod_{j=1}^n ~
{\theta_1(a_j | \tau) \over \theta_1(b_j | \tau)}  \ ,}
where
\eqn\abdefn{\eqalign{ a_j ~&=~ \big(1 - \coeff{1}{k+n+1} \big)~
\mu_j~-~\mu_{j-1} ~+~ \big(1 - \coeff{j}{k+n+1} \big)~ \nu \ , \cr
b_j ~&=~  - \coeff{1}{k+n+1} ~\mu_j~-~ \coeff{j}{k+n+1} ~ \nu \ ,}}
with the convention that $\mu_0 = \mu_n \equiv 0$. The parameters,
$\mu_j$ are defined by writing $\mu \cdot j(z) = \sum_j
\sqrt{\coeff{k+n} {k+n+1}}~\mu_j
\gamma_j \cdot \partial \sigma(z)$.  Recall that we may write
$\gamma_j = e_j - e_{j+1}$, $j =1, \dots, n-1$.  The Weyl group
of $SU(n)$ is the permutation group on $n$ objects, acting
in the obvious manner on the $e_j$.  From this it is trivial to
determine the Weyl action on the $\mu_j$, and hence obtain the
function ${\cal F}_s$ from ${\cal F}$.

For example, taking $n=2$, one obtains:
\eqn\Ftwo{\eqalign{{\cal F}(q,y,z) ~=~  y^{-1} z^k ~
\prod_{p=1}^\infty~ \Bigg\{ ~ & {{(1 - q^{p-1} y^{-(k+2)}
z^{(k+2)})~(1 - q^p y^{(k+2)} z^{-(k+2)}) } \over
{(1 - q^{p-1} y^{-1} z) ~(1 - q^p y z^{-1}) } }
\cr & {{(1 - q^{p-1} y^{(k+3)} z^{(k+1)}) ~(1 - q^p y^{-(k+3)}
z^{-(k+1)}) }
\over  {(1 - q^{p-1} z^2) ~(1 - q^p z^{-2}) } } ~ \Bigg\}  \ . }}
where $y=exp\big[-{i { \mu} \over \sqrt{(k+2)(k+3)}}\big]$ and
$z=exp[ -{i\nu \over k+3}]$.  One can immediately see that this
function is singular at $z=1$, and thus it cannot be a
character of a unitary coset model.  However, to Weyl symmetrize,
one simply replaces $\mu$ by $-\mu$, and obtains:
\eqn\Fstwo{{\cal F}_s(q,y,z) ~=~ {\cal F}(q,y,z) ~+~
{\cal F}(q,y^{-1},z) ~=~ {\cal F}(q,y,z)
{}~+~ {\cal F}(q,y,z^{-1}) \ .}
This function is regular at $z=1$, and extensive expansion using
${\it Mathematica}^{TM}$ confirms that it generates the proper
characters of the factors in \tensor.

\subsec{Decomposing the refined elliptic character}

To complete the process of isolating the component characters of
the model we need to extract the coefficient of terms of
the form \phaseform\ in the Weyl symmetrized form of \Fform.
To do this, one needs to expand the theta functions
in the denominator of \Fform\ using the
identity \ref\PvN{C.B.~Thorn, ``String Field Theory,'' \prpt{175}
(1989) 1; P.~Bouwknegt, A.~Ceresole, J.G.~McCarthy and
P.~van~Nieuwenhuizen,  \prd{39} (1989) 2971; D.Z.~Freedman
and K.~Pilch \ijmp{A4} (1989) 5553.}:
\eqn\invtheta{\eqalign{ {1 \over \theta_1(\nu |\tau)} ~&=~
\Big[~\big(e^{i\pi \nu} - e^{-i\pi \nu} \big) ~\prod_{p=1}^\infty
{}~\big(1 - q^p e^{2 \pi i \nu} \big) ~\big(1 - q^p e^{- 2 \pi i \nu}
\big) ~\Big]^{-1} \cr
{}~&=~ i~q^{-{1 \over8}}~ \Big[ \prod_{n=1}^\infty ~(1 - q^n)^{-3}~
\Big]~ \times \cr
&\qquad \qquad \qquad \sum_{\ell = -\infty}^\infty ~
\sum_{p=0}^\infty ~(-1)^p~ e^{2 \pi i (\ell - \half) \nu} ~
q^{\half(p \pm \ell + \half)^2 - \half(\ell - \half)^2} \ .}}
In this formula one can choose the $\pm$ sign in any manner one
pleases because of the identity:
\eqn\stdid{\sum_{p=0}^{2m-1} ~(-1)^p~ q^{\half(p - m + \half)^2 -
\half(m - \half)^2} ~\equiv~ 0 \ .}

To give the expressions for the branching functions of ${\cal M}_1$
as they emerge from the elliptic genus, we need to introduce some
notation. We will need another basis for the roots of $SU(n+1)$:
\eqn\rtbasis{ \bar \alpha_j ~\equiv~ e_{n+1-j} ~-~ e_{n+1}
\ , \quad j = 1, \dots, n \ ;}
along with the  corresponding dual weight basis (satisfying
$\bar \lambda_i \cdot \bar \alpha_j = \delta_{ij}$):
\eqn\wtbasis{\bar \lambda_j ~\equiv~ e_{n+1 -j} ~-~ {1 \over n+1}
(e_1 + \dots + e_{n+1})   \ , \quad j = 1, \dots, n \ .}
Given any vector, $\zeta$, define vector and scalar
projections, $\zeta_0$ and $\hat \zeta$, via:
\eqn\bitsdefn{\eqalign{\zeta ~&=~ \sum_{j=1}^n ~\zeta_j~\bar
\alpha_j \ ; \qquad
\hat \zeta ~=~ 2 (\tilde \rho - \rho) \cdot \zeta ~=~ (n+1)~
\sum_{j=1}^n ~\zeta_j  \ ; \cr
\zeta_0 ~&=~ \zeta ~-~ \coeff{2 \hat \zeta}{ n(n+1)} ~
(\tilde \rho - \rho) ~=~\sum_{j=1}^n ~\zeta_j ~\big[ e_{n+1-j} ~-~
{1 \over n}   (e_1 + \dots + e_{n}) ~\big] \ .}}
Recall that $\tilde \rho$ and $\rho$ are the Weyl vectors of
$SU(n+1)$ and $SU(n)$ respectively, and that $\tilde \rho - \rho$
defines the $U(1)$ direction in ${\cal M}_1$.  Thus
$\zeta_0$ and $\hat \zeta$ are the components
of $\zeta$ parallel and perpendicular to the $U(1)$.
Introduce two vectors:
\eqn\uvdefn{\eqalign{v ~&\equiv~ \coeff{1}{k+n+1}~
\Big[~ \nu ~\tilde \rho ~+~ \sum_{j=1}^{n-1} ~ \mu_j~ \bar
\lambda_j~\Big] \ ,\cr
u ~&\equiv~ \coeff{2 \nu}{n+1}~( \tilde \rho ~-~ \rho) ~-~ v ~-~
\sum_{j=1}^{n-1} ~ \mu_j~  (\bar \lambda_{j+1} - \bar \lambda_j) \
.}}
The whole point of these vectors is that
$\bar \alpha_j \cdot u ~=~ a_j$ and $\bar \alpha_j \cdot v ~=~-b_j$,
where $a_j$ and $b_j$ are given by \abdefn.  Finally,
introduce a vector $\xi$ defined by
\eqn\psdefn{\xi ~\equiv~ \sum_{j=1}^{n} ~ p_j~ \bar \lambda_j \ ,
\quad {\rm with} \quad \ p_j \geq 0 \ .}
This vector will generate all the sums over $p_j \geq 0$ when
we invert the denominators of \Fform\ using \invtheta.

The function, ${\cal F}(q,\mu,\nu)$, can then expanded
according to:
\eqn\Fexpand{\eqalign{{\cal F}(q,\mu,\nu) ~=~ {1 \over
\eta(q)^n}   \sum_{\beta \in \Gamma}
{}~\sum_{\lambda \in \Gamma^*} ~ & e^{i \pi (\hat u + \hat v
+ {1 \over (n+1)} \hat \beta  )}~ e^{- {2 \pi i \over
(k+n+1)(n+1)} (\hat u + \hat v) \hat \lambda } \cr
& e^{- 2 \pi i (u+v) \cdot ( \beta + {\lambda \over k+n+1}) }~
q^{\Delta(\beta,\lambda)} ~ G_{\beta, \lambda} (q) \ ,}}
where $\Gamma$ is the root lattice of $SU(n+1)$, $\Gamma^*$
is the weight lattice of $SU(n+1)$, and
\eqn\Deltadefn{\eqalign{ \Delta(\beta,\lambda) ~\equiv&~
\coeff{1}{2(k+n)(k+n+1)} \big[~\lambda_0 ~+~ (k+n+1) \beta_0 ~
\big]^2 \cr
\quad ~&+~
\coeff{1}{2kn(n+1)^2 (k+n+1)} \big[~(n+1) \hat\lambda~+~
(k+n+1) \hat \beta \cr
& -~ \shalf n(n+1)(k+n+1) ~\big]^2  \ .}}
The quadratic form, $\Delta(\beta,\lambda)$, is the energy in
${\cal M}_2 \times U(1)$ of the momentum state labelled by
$\beta + {\lambda \over k+n+1}$.
After a considerable amount of work, the functions
$G_{\beta, \lambda}$ can be written:
\eqn\Gdefn{G_{\beta, \lambda} ~\equiv~
{1 \over \eta(q)^{2n}} ~ \sum_{\eta \in \Gamma^*} ~\sum_{\xi} ~
e^{i \pi (\hat \xi + \hat \eta)}~
q^{\Omega(\beta,\lambda; \xi, \eta)} \ ,}
where $\xi$ is defined by \psdefn\ and the sum is, of course, only
over $p_j \ge 0$.   Because of the conditional convergence of the
double sum in \invtheta, one must perform the sum over $\xi$ before
performing the sum over $\eta$ in \Gdefn.  One must
also be very careful in performing any reordering of this sum, and
in shifting summation variables.
The quadratic form $\Omega(\beta,\lambda;\xi,\eta)$ is defined by:
\eqn\Omdefn{\eqalign{\Omega(\beta,\lambda; \xi, \eta) ~\equiv&~
\coeff{1}{2(k+n+1)} \big[~(k+n+1) (\xi + w_c(\eta)) ~+~
w_c(\lambda) ~ \big]^2 \cr
\quad ~&-~ \coeff{1}{2(k+n)} \big[~(k+n) (\xi_0 +
\eta_0) ~+~  (\beta_0 + \lambda_0) ~ \big]^2 \cr
\quad ~&-~
\coeff{1}{2kn(n+1)} \big[~k (\hat \xi + \hat{\eta}) ~+~
(\hat\lambda ~+~ \hat \beta) ~-~ \shalf n(n+1) ~\big]^2
\ .}}
The function, $w_c$, is a cyclic Weyl rotation of $SU(n+1)$
that takes $e_1 \to e_2 \to \dots \to e_{n+1} \to e_1$.  Note that
one does not sum over $w_c$ in any manner, it is simply used to
transform the vectors $\lambda$ and $\eta$ in \Omdefn.

Finally, to get the branching functions,
$G^{(s)}_{\beta, \lambda}$, of ${\cal M}_1$,
we must Weyl symmetrize.  That is,
\eqn\Gsdefn{G^{(s)}_{\beta, \lambda} ~\equiv~
\sum_{w \in W(SU(n))} ~ G_{w(\beta), w(\lambda)}  \ .}

\subsec{Branching functions of ${SU_k(n+1) \over SU_k (n)
\times U(1)}$}

The direct way of obtaining the branching functions of ${\cal M}_1$
is from the Weyl-Kac character formula.   We will now do this so as
to obtain expressions that can be compared with those for
$G^{(s)}_{\beta, \lambda}$.  One expands the
appropriate parts of the denominator of the character formula
using \invtheta\ and then factors out the characters of
$SU_k(n) \times U(1)$ from the resulting expression. The
coefficient functions of the $SU_k(n) \times U(1)$ characters
are then the branching functions.
If $\Lambda$ is a highest weight of $SU_k(n+1)$, and
$\chi$ is a weight of $SU_k(n)  \times U(1)$, then the
branching functions are only non-zero if $\chi
= \Lambda + \beta$ for some root $\beta$ of $SU(n+1)$,
and then one has
\eqn\branch{{b^\Lambda}_\chi ~=~
{1 \over \eta(q)^{2n}}~ \sum_{w \in W(SU(n+1))}~\sum_{\gamma \in
\Gamma}~\sum_\xi ~\epsilon(w) ~(-1)^{\hat \xi} ~q^{\widetilde
\Omega_w (\Lambda,\chi; \xi,\gamma)} \ .}
The sum over $\xi$ is exactly as above, and $\epsilon(w)$ is,
as usual, the determinant of $w$.  Note that unlike
above, the sum over $w$ is over the Weyl group of
$SU(n+1)$ and not just that of $SU(n)$.
The quadratic form $\widetilde \Omega_w(\Lambda,\chi;
\xi,\gamma)$ is closely related to $\Omega$ in \Omdefn:
\eqn\Omtilde{\eqalign{\widetilde \Omega_w(\Lambda,\chi;
\xi, \gamma) ~\equiv&~ \coeff{1}{2(k+n+1)} \big[~(k+n+1) (\xi +
\gamma)~+~ w(\Lambda + \tilde \rho) ~ \big]^2 \cr
\quad ~&-~ \coeff{1}{2(k+n)} \big[~(k+n) \xi_0  ~+~
 (\chi_0 + \rho) ~ \big]^2 \cr
\quad ~&-~ \coeff{1}{2kn(n+1)} \big[~k \hat \xi  ~+~
\hat {\chi} ~\big]\ .}}
These branching functions have the following symmetries:
\eqn\brsymmone{{b^\Lambda}_\chi ~=~ {b^\Lambda}_{\chi + (k+n)
\beta_0} ~=~ {b^\Lambda}_{\chi + 2 k(\tilde \rho - \rho)} \ ,}
where $\beta_0$ is a root of $SU(n)$.  There are also the spectral
flow identifications: The vector $\lambda_{n+1} = \coeff{2}{n+1}
(\tilde \rho - \rho)$  is a weight of $SU(n+1)$.  For any given
$\Lambda$, there is a root, $\beta$, of $SU(n+1)$, and an element,
$w^\prime$, of  the Weyl group of $SU(n+1)$, such that the vector,
$\Lambda^\prime$, defined by:
\eqn\lampdefn{\Lambda^\prime  ~\equiv~ w^\prime (\Lambda +
\tilde \rho) ~+~ (k+n+1)(\lambda_{n+1} + \beta) ~-~  \tilde \rho\ ,}
is once again an affine label of $SU_k(n+1)$.
For such vectors $\Lambda$ and $\Lambda^\prime$ one has:
\eqn\specflw{ {b^\Lambda}_\chi ~=~ \epsilon(w^\prime)~
{b^{\Lambda^\prime}}_{\chi  + k \lambda_{n+1}} \ .}

{}From the Weyl-Kac character formula, it is natural to extend the
label $\Lambda$, of the branching function, to any vector
$\Lambda^\prime$ on the weight lattice of $SU(n+1)$.  That is,
one takes ${b^{\Lambda^\prime}}_\chi = 0$ if
$(\Lambda^\prime +\tilde \rho) \cdot \beta \equiv 0$ mod $k+n+1$ for
any root $\beta$ of $SU(n+1)$.  For any other weight,
$\Lambda^\prime$,
there is a root $\beta$, and a Weyl rotation, $w$, such that
$\Lambda = w(\Lambda^\prime +  \tilde \rho) - \tilde \rho + (k+n+1)
\beta$ is a highest weight of affine $SU_k(n+1)$.  We therefore take:
\eqn\symmtwo{ {b^{\Lambda^\prime}}_\chi ~=~ \epsilon(w)~
{b^\Lambda}_\chi
\qquad {\rm where} \quad \Lambda ~=~ w(\Lambda^\prime +
\tilde \rho) - \tilde \rho + (k+n+1) \beta \ .}

{}From our general arguments about the properties of the refined
elliptic genus, the foregoing branching functions are related to the
functions  $G^{(s)}_{\beta, \lambda}$ by:
\eqn\Gbreln{{b^\Lambda}_\chi ~=~  G^{(s)}_{\beta , \lambda}
\quad {\rm with} \quad \lambda = w(\Lambda + \tilde \rho) \ ;
\quad \beta + \lambda = \chi + \tilde \rho \ .}
The Weyl element $w$ in this relation can be chosen at will.

One can easily see that the following replacements:
\eqn\repl{\xi \to \xi - \eta \ ; \quad w_c(\eta) - \eta \to
\gamma \ ;  \quad \lambda \to w(\Lambda + \tilde \rho) \ ;
\quad \beta \to \chi - [w(\Lambda + \tilde \rho) - \tilde \rho]\ ,}
transform $\Omega$ of \Omdefn\ directly into $\tilde
\Omega_{w^\prime}$ with $w^\prime = w_c w$.  The problem with going
further and directly establishing the identity \Gbreln, independently
of the elliptic genus, is that the sums in \Gdefn\ and \branch\
are conditionally convergent, and thus must be handled with
considerable care.   We have thus only been able to prove \Gbreln\
directly for $n=1$ and $n=2$, and based upon this, we believe that
a general direct proof will require breaking
sums over the root or weight lattice into many sums over different
cones
on the lattice and then making extensive rearrangements and
use of the identity \stdid.
Since we have the general argument based on the elliptic genus,
the direct proof for $n=1$ and $n=2$, as well as extensive checks
using $Mathematica^{TM}$, we have not pursued a general direct
proof any further.

\subsec{A simple example}

\nref\KacPet{V.G.~Kac and D.~Peterson, \advm{53} (1984) 125.}
\nref\GepQ{D.~Gepner and Z.~Qiu \nup{285}  (1987) 423;
D.~Gepner, \nup{296} (1988) 757.}

For $n=1$ the forgoing functions, $G^{(s)}$, are labelled with
by two integers (a root and a weight of $SU(2)$) and are the
branching functions of $SU(2)/U(1)$.  That is, we will recover
the string functions, $c^\ell_m$, that are the partition functions
for parafermions.  Taking $\xi = {p \over 2} (e_1 - e_2)$,
$\eta = - {n \over 2} (e_1 - e_2)$, $\lambda = -{(\ell + 1) \over 2}
(e_1 - e_2)$ and $\beta = (s+1) (e_1 - e_2)$ in \Gdefn,
one arrives at:
\eqn\Gneqlone{G^{(s)}_{s, \ell}(\tau) ~=~{1 \over \eta(q)^2}~
\sum_{n= -\infty}^\infty~ \sum_{p=0}^\infty~ (-1)^{n+p}~
q^{{1 \over 4(k+2)} [(k+2)(n+p) + (\ell+1) ]^2~-~
{1 \over 4k} [k(p-n) + 2s - \ell ]^2 } \ .}
The standard form of the string functions of parafermionic models
is \refs{\KacPet,\GepQ}:
\eqn\hecke{\eqalign{ c^\ell_m(\tau) ~=~ \sum_{-|x| < y \leq |x|}~
sign(x) ~ & q^{(k+2) x^2 - ky^2} \ ; \cr
{\rm with\ } & (x,y) {\rm \ or\ } (\shalf - x,\shalf + y)
\in \big(\coeff{\ell+1}{ 2(k+2)}, \coeff{m}{2k} \big) +
\ZZ^2  \ .}}
An equivalent form for the string function can also be obtained
from \branch.  Taking $\xi = {p \over 2} (e_1 - e_2)$,
$\gamma = n (e_1 - e_2)$, $\Lambda = {\ell \over 2}
(e_1 - e_2)$ and $\chi = {m \over 2} (e_1 - e_2)$, one obtains:
\eqn\stringfn{c^\ell_m(\tau) ~=~{1 \over \eta(q)^2}~
\sum_{\epsilon = \pm 1}~ \sum_{n= -\infty}^\infty~
\sum_{p=0}^\infty~ \epsilon~(-1)^{p}~  q^{{1 \over 4(k+2)}
[(k+2)(2n+p) + \epsilon (\ell+1) ] ~-~ {1 \over 4k}
[kp + m ]^2 } \ ,}
with the selection rule $m \equiv \ell$ mod $2$.

It turns out to be a little involved to establish directly that
these three forms for $c^\ell_m$ of are equivalent.
The easiest is to show that $G^{(s)}_{s, \ell}(\tau)
= c^\ell_{2s-\ell}(\tau)$, where the latter is given by \hecke.
One simply has to parametrize the sums in \hecke\ over the
four sectors of the $(x,y)$ plane, make modest use of \stdid, and
then regroup the sums into the form of \Gneqlone.  The equivalence
with \stringfn\ requires that one start by first breaking the sum
into $n \geq 0$ and $n < 0$, and then breaking one of the two
resulting sums into sums with $p \geq n$ and $p < n$, while the other
sum is broken into sums with $p > n$ and $p \leq n$.
One then appropriately relabels the summation variables, makes use of
\stdid, and regroups the terms.  The result is \stringfn.

Thus one sees that the refined elliptic genus provides us with
precisely the proper branching functions.

\newsec{Fermionic screening}

\nref\FMS{D.\ Friedan, E.\ Martinec and S.\ Shenker, \nup271(1986)
93.}
\nref\WLNW{W.~Lerche and N.P.~Warner, ``On the Algebraic Structure
of Gravitational Descendants in CP(n--1) Coset Models,'' USC and
CERN preprints CERN-TH7442/94, USC-94/014, hep-th/9409069.}

Before making some general comments about our results, there is one
minor, and perhaps interesting, loose end that needs to be addressed.

So far we have accounted for all of the $D$-type screening that is
involved in the $N=2$ supersymmetric Coulomb gas description, but we
have, as yet, said very little about the fermionic screening. This is
most easily understood by looking at the simplest model, with $n=1$.
This model is based upon $SU_k(2)$, which has a Kac-Wakimoto
realization in terms of a bosonic $\beta$--$\gamma$, or superghost,
system and a single free boson. There is a single screening current
for this model, and it can be written as a product of $\beta(z)$ and
a bosonic vertex operator. To get to the Coulomb gas realization of
the $N=2$ model, one first bosonizes the superghost system according
to \FMS:
\eqn\bgbosons{\eqalign{
\beta \ &=\ (\del\xi)~e^{-\phi}\ =\ i(\del\chi)~e^{i\chi-\phi}\cr
\gamma \ &=\ \ \ \eta~ e^\phi\ \ \ =\ e^{-i\chi+\phi} \ .}}
Next, one tensors in a new $U(1)$, and factors out the appropriate
diagonal $U(1)$ factor to arrive at the coset model
$SU(2) \times U(1)/U(1)$.  In this process, the $SU_k(2)$ screening
current maps directly onto the $D$-type screener of the $N=2$
supersymmetric model.

The necessity of having a fermionic screener creeps in at the point
where one bosonizes the $\beta$--$\gamma$ system. To recover the
Hilbert space of the superghosts from the Hilbert space of the
$\xi$--$\eta$, and $\phi$ system, or from the $\phi$--$\chi$ Fock
space, of \bgbosons, one has to fix the momenta $p_\phi-p_\chi$ and
exclude all states involving the zero mode, $\xi_0$ \FMS. An
equivalent way of accomplishing the same thing is that one can allow
states with $p_\phi-p_\chi\geq0$, and then compute the cohomology of
the fermionic charge $Q=\oint\eta$ \BMP. Once again, if one
translates this across to the $N=2$ supersymmetric Coulomb gas
language, one finds that this is precisely what is done by the
fermionic screener. If one does not employ the fermionic screeners,
one obtains infinitely many copies of the Hilbert space of the $N=2$
supersymmetric model. These infinitely many copies are related by
shifts in the momenta $p_\phi-p_\chi$.

The foregoing observations generalize in a fairly obvious way to the
$N=2$ supersymmetric Coulomb gas description of \coset\ for arbitrary
$n$. Indeed in \WLNW\ it was shown how the copies of the physical
Hilbert space can be reinterpreted as gravitational descendants of
the matter sector. In terms of the characters derived in the last
section, the fermionic screening can be seen rather explicitly as
being responsible for the sums over the vector $\xi$ in \Gdefn\ and
\branch. As a consequence of this we see that the elliptic genus has
implicitly taken care of this fermionic screening as well. The moral
reason for why this happens is probably related to the fact that the
elliptic genus originates from the \LG formulation which can be
intrinsically expressed in terms of superghosts as in sect. 3.3. It
is only when the superghosts are bosonized that one needs to worry
about the fermionic screeners explicitly.

The \LG formulation, along with the work of \WLNW, suggests another
interesting possibility for the fermionic screening charges.
Specifically, they can also be incorporated as parts of the
supercharges, as in \potscreen.  The immediately apparent obstacle
to doing this in the Coulomb gas language is that by definition,
the screening charges commute with the chiral algebra, whereas the
supercharges have non-zero $U(1)$ charge.  To rectify this, one
makes a very simple change to the fermionic screeners.  Introduce
the operators:
\eqn\supermods{\eqalign{
\widetilde G^-(z,\bar z)  \ &= \  \sum_{i = 1}^n ~\bar \psi_i ~
e^{{i  \sqrt{k+n+1}} ~ \bar \phi_i(z,\bar z)} \ , \cr
\widetilde G^+(z,\bar z)  \ &= \  \sum_{i = 1}^n ~
(\psi_i ~-~ \psi_{i-1})~e^{{i  \sqrt{k+n+1}} ~ (\phi_i(z, \bar z)  -
\phi_{i-1}(z, \bar z))} \ . } }
At first sight, these operators appear to be nothing other than sums
of the fermionic screeners in \newscreener. The crucial difference is
that the bosons in the vertex operator part are to be taken as the
complete boson, {\it i.e.} as a function of $z$ and $\bar z$, and not
just the holomorphic, or left-moving, part. The effect of doing this
is to give the operators, $\widetilde G^\pm (z,\bar z)$ a
right-moving $U(1)$ charge of $\pm 1$. The idea is to now view
$\widetilde G^\mp(z,\bar z) $ as anti-holomorphic components of a
conserved current whose holomorphic components are $G^\pm(z)$. (There
will be similar holomorphic components to the anti-holomorphic, or
right moving supercurrents $\bar G^\mp(\bar z)$.) The motivation for
doing this comes from the \LG formulation and the corresponding
corrections to the supercurrent due to the presence of a non-trivial
superpotential. In the Coulomb gas language, the corrections to the
supercharge become essential if one imagines perturbing the model by
the conformal weight $(1,1)$ operators $\sum_i S_i(z) \bar S_i(\bar
z)$, where the $S_i$ are the fermionic screening currents. In
correlators, such perturbations yield the proper screening
prescriptions in the conformal blocks. Hence if one does not perform
the fermionic screening, then the supercharges receive the foregoing
corrections. This is analogous to viewing the \LG potential as a
perturbation of the free theory.

One might naturally ask what one learns from this apparently somewhat
perverse perspective. First, the \LG potential does generate the
foregoing modifications to the conserved supercurrents, and so
connecting these modifications with screening currents yields a
rather better understanding of the infra-red limit of the
renormalization group flow of the \LG theory. On a much more
practical level, it was very much part of the original thinking in
the \LG program \LGboth\ that the \LG potential should encode the
structure of the modular invariant partition function. Thus one would
hope that the same is true for the foregoing modifications of the
supercurrents. In particular, if one looks at the right-moving vertex
operator parts of \supermods\ then these will map one between
different representations of the extended $N=2$ super-chiral algebra,
whereas the left-moving screening charge part will map into
effectively equivalent representations. Some preliminary
investigations for the simplest model indicate that this is true. If
one considers the model \coset\ with $n=2$, then the partition
function contains combinations of string functions and $U(1)$
characters for both the left and right moving sector. The foregoing
vertex operator shifts the $N=2$, $U(1)$ charge by one, {\it and}
shifts the $m$ quantum number on $c^\ell_m$ by two units. This
suggests that in the modular invariant partition function, a given
left-moving character $c^\ell_m$ will be paired with all the right
moving characters $c^\ell_{m+2p}$ for all $p$. It also indicates a
particular correlation of these quantum numbers and $N=2$ $U(1)$
charges. This is in fact what one finds when one decomposes the
partition function into its component parts. This particular example
is a little trivial since we have deduced something that the we could
have easily inferred from the presence of the supercharge itself.
However, for models with more than one \LG variable the foregoing
procedure will generate more than one vertex operator, and we will
get more than one set of correlations between left-moving
and right-moving characters. The situation is a little reminscent of
the lattice structure that underlies non-trival modular invariants of
affine Lie algebras \ref\NWindex{N.P.~Warner, \cmp{130} (1990) 205.}.

Thus we suspect that the \LG potential, and its Coulomb gas
concomitants, contain the information about how left-moving and
right-moving representations are locked together, and that the
foregoing might provide a method of explicitly extracting this
information.

\newsec{Final Comments}

\nref\MNW{Z.~Maassarani, D.~Nemeschansky and N.P. Warner, \nup{393}
(1993) 523.}
\nref\offcrit{D.~Nemeschansky and N.P. Warner, \nup{413} (1994) 629.}

We have shown in some considerable detail how the various
formulations of the $N=2$ super-$W$ minimal models are interrelated
and have shown how the elliptic genus and the \LG potential can be
used to get very detailed information about the partition functions
of these models. There also remains the interesting question as to
how to decode from the \LG potential the content of the modular
invariant.  One would also like to know if one can get information
from the elliptic genus about the complete partition function of the
model, and not just about the characters above the Ramond ground
states.

There are also natural questions about the underlying exactly
solvable lattice models. Given that these models have now been
constructed \refs{\MNW,\offcrit}, one might hope to adapt some of
the topological index results to the lattice model. One possible
hope might be to extract the elliptic index from the lattice
formulation without having to resort to detailed computations
involving Bethe Ansatz or the corner transfer matrix. The fact that
the free energy of these lattice models vanishes for topological
reasons \offcrit\ gives us hope that the lattice models may contain
other pieces of topological information.

{}From the point of view of the field theory alone, we think it
remarkable that so much of the structure of the theory can be deduced
from the \LG potential alone.  It compelling to see if yet more
information can be obtained about other related, and perhaps even
massive, models using \LG methods.

\appendix{A}{Determining $W_3$ in the \LG formulation}

In this appendix we will consider the \LG formulation of the model
\coset, with $n=2$, and we will obtain the $W_3$-current by writing
down the most general Ansatz in superspace, imposing chirality and
using the \LG equations of motion. The current we are looking for has
dimension two and therefore the top component of the current has
dimension three. The possible terms in the ansatz can be reduced by
realizing that current has to be neutral. This means that it contains
an equal number of $\Phi_i^+$ and $\Phi^-_i$ fields. The most general
Ansatz contains eighteen terms and has the form:
\eqn\ansa {\eqalign{ {\cal W} \ = \ & a_1 D^+ \Phi_1^+ D^-\Phi_1^-
D^+ \Phi_2^+ D^-\Phi_2^-+ a_2 \Phi^+_1 \del \Phi^-_1 D^+ \Phi_2^+
D^-\Phi_2^-+\cr
&  a_3 \Phi^+_1 \del \Phi^-_1 D^+ \Phi_1^+ D^-\Phi_1^- +
 a_4 \Phi^+_2 \del \Phi^-_2 D^+ \Phi_2^+ D^-\Phi_2^-+\cr
& a_5 \Phi^+_2 \del \Phi^-_2 D^+ \Phi_1^+ D^-\Phi_1^-
+a_6 \Phi^+_1 \del \Phi^-_2 D^+ \Phi_2^+ D^-\Phi_1^-+\cr
& a_7 \Phi^+_2 \del \Phi^-_1 D^+ \Phi_1^+ D^-\Phi_2^-+
a_8 \del \Phi^+_2\del\Phi^-_2+a_9 \del \Phi^+_1\del\Phi^-_1+\cr
& a_{10} D^+\Phi^+_1\del D^-\Phi_1^- + a_{11} D^-\Phi^-_1\del
D^+\Phi_1^+ +
a_{12} D^+\Phi^+_2\del D^-\Phi_2^- +  \cr
&a_{13} D^-\Phi^-_2\del D^+\Phi_2^+ +
a_{14}\Phi^+_1 \del^2\Phi_1^-+a_{15}\Phi^+_2 \del^2\Phi_2^- +\cr
& a_{16}\Phi^+_1\Phi^+_1\del\Phi^-_1\del\Phi^-_1+
 a_{17}\Phi^+_2\Phi^+_2\del\Phi^-_2\del\Phi^-_2+
a_{18}\Phi^+_1\Phi^+_2\del\Phi^-_1\del\Phi^-_2 \ , \cr }}
where $a_i$ are unknown coefficients.  Most of these coefficients
are determined by requiring that $W$ satisfy:
\eqn\conserv{\bar D^- {\cal W} \ = \ 0 \ . }
In simplifying the expression that results from \conserv\ one
uses the \LG equations of motion:
\eqn\eqmo{\bar D^-D^- \Phi^-_i \ = \ \coeff{1}{2} { \del W \over \del
\Phi^+_i } \ ,}
along with the fact that the superpotential $W(\Phi^+_1,\Phi^-_1)$
has a very specific form. First, one needs to use the fact that $W$
is quasihomogeneous, and hence:
\eqn\quasihom{ W \ = \ {1 \over k+3} ~ \bigg[ \Phi^+_1~{\del W \over
\del \Phi^+_1} ~+~ 2~\Phi^+_2~{\del W \over \del
\Phi^+_2} \bigg] \ .}
Secondly, the fact that the potential comes from a coset model
determines its form uniquely.  Indeed, the exact form is given
implicitly by \LVW:
\eqn\susypot{ W \ = \ \xi^{k+3}_1 ~+~ \xi^{k+3}_2 \ , }
where
\eqn\defxii{ \Phi_1^+ \ = \ \xi_1 + \xi_2 \quad {\rm and} \quad
\Phi^+_2\ = \  \xi_1 ~\xi_2 \ .}
This form of the potential is uniquely characterized (up to
trivial scaling) by the equation $\partial^2 W/\partial \xi_1
\partial \xi_2 = 0$, which may be rewritten as:
\eqn\fixesW{{\del^2 W \over (\del {\Phi^+_1})^2 } ~+~ \Phi^+_1~
{ \partial^2 W \over \del  \Phi^+_1\Phi^+_2} ~+~  \Phi^+_2~
{\del^2 W  \over (\del {\Phi^+_2})^2\ } ~+~ {\del W \over \del
{\Phi^+_2} } ~=~ 0\ .}
This, along with quasihomogeneity, implies numerous relationships
between partial derivatives of $W$.  In particular, one finds:
\eqn\relpart{\eqalign{ {\del^2 W \over (\del {\Phi^+_1})^2 } \ = & \
-{ k+2\over  k+1 }~\Phi^+_1~  { \partial^2 W \over \del \Phi^+_1
\del \Phi^+_2}~-~ {k+3 \over k+1} ~ \Phi^+_2~  {\del^2 W \over (\del
{\Phi^+_2})^2\ } \cr
{\del W \over \del \Phi^+_1 } \ = & \  {\Phi^+_1\over { k+2}}
{ \partial^2 W \over (\del \Phi^+_1)^2 } ~+~ {2 \Phi^+_2 \over
{ k+2} }~{ \partial^2 W \over \del \Phi^+_1 \del  \Phi^+_2}\cr
{\del W \over \del \Phi^+_2 } \ = & \ {\Phi^+_1\over { k+1}}{
\partial^2 W \over \del \Phi^+_1\Phi^+_2}~ + ~ {2\Phi^+_2\over {
k+1}} { \partial^2 W \over (\del {\Phi^+_2})^2} \ .
\cr }}
Conversely, it is only when these relations \relpart\ are
satisfied that one can find a non-trivial solution to \conserv\
for some choice of the coefficients $a_i$.
After some algebra   we found the following solution:
\eqn\solution{\eqalign{ {\cal W } \ =  \ & b_1 ~{\cal J }^2 ~+~
b_2 ~ \del {\cal J} ~+~ b_3~ (D^+ D^- - D^-D^+ ) {\cal J} \cr
& +~ b_4~\Bigl ( -\coeff{1}{2}\hat{\cal   J}^2 ~+~
\coeff{i}{\sqrt{2(k+2)(k+3)}}~ \del \hat{\cal J}  ~-~
\coeff{1}{2 (k+2)} \Phi^-_1\Phi^+_2D^+\del\Phi^+_1D^-\Phi^-_2 \cr
& ~-~ \coeff{i \sqrt{k+3}}{ 2 (k+2)}~
\Phi^+_1\del\Phi^-_1D^-\del\Phi^-_2D^+\Phi^+_1\Bigr ) \ , }}
where
\eqn\defhatj{\hat{\cal J} \ \equiv\  \coeff{i}{\sqrt{2(k+3)(k+2)}}~
\Bigl ( \Phi^+_1 \del \Phi^-_1 - \coeff{1}{2} (k+2)
D^+\Phi^+_1D^-\Phi^-_1+\coeff{1}{2}(k+3)D^+\Phi^+_2D^-\Phi^-_2 \Bigr
) }
The terms ${\cal J}^2 $, $\del {\cal J} $ and $ (D^+D^--D^-D^+){\cal
J}$ correspond to trivial solutions since ${\cal J}$ is the only
dimension one supercurrent that is conserved ({\it i.e.} satisfies
the chirality condition \conserv). We can fix the coefficients in
${\cal W}$ up to an overall normalization by demanding ${\cal W}$
has the appropriate operator product expansion with ${\cal J}$.
This gives:
\eqn\fianlww{\eqalign{ {\cal W}  = & \coeff{k(k+5)}{2(k+2)(k+3)}
(D^+D^--D^-D^+) {\cal J} ~-~ \coeff{(k+5)}{ k+2}~:{\cal J}^2 : \cr
& - ~ \coeff{(5k-3)}{k+3} \Bigl(- \coeff{1}{2} \hat{\cal J}^2 ~+~
\coeff{i}{\sqrt{2(k+2)(k+3)}} ~\del \hat {\cal J} \cr
&-~\coeff{1}{2 (k+2) } ~\Phi^-_1\Phi^+_2D^+\del\Phi^+_1D^-\Phi^-_2
{}~-~ \coeff{i \sqrt {k+3}}{ 2(k+2)}~\Phi^+_1\del\Phi^-_1D^-
\del\Phi^-_2D^+\Phi^+_1 \Bigr )}}
Note that \fianlww\ has a form that is identical to the one we
obtained from the Drinfel'd-Sokolov reduction \callw. The only
apparent difference is in the definitions of the currents
${\cal J}$ and $\hat {\cal J}$.  However, these currents
can be mapped onto one another using the translation table
\transl.

\vfill
\eject
\listrefs

\vfill
\eject
\end